\documentclass{jfm}
\usepackage{graphicx}
\usepackage{newtxtext}
\usepackage{newtxmath}
\usepackage{natbib}
\usepackage{hyperref}
\usepackage{indentfirst}
\usepackage{amsmath}
\usepackage{bm}
\usepackage{float}
\usepackage[width=\textwidth,justification=justified]{caption}
\hypersetup{
    colorlinks = true,
    urlcolor   = blue,
    citecolor  = black,
}

\newcommand{\RomanNumeralCaps}[1]
\linenumbers

\title{Linear stability theory and molecular simulations of nanofilm dewetting with disjoining pressure, strong liquid-solid slip, and thermal fluctuations}

\author{Yixin Zhang\aff{1}\corresp{\email{Y.Zhang-11@utwente}}}
\affiliation{\aff{1}Physics of Fluids Group, Max Planck Center Twente for Complex Fluid Dynamics and J. M. Burgers Centre for Fluid Dynamics, University of Twente, P.O. Box 217, 7500 AE Enschede, The Netherlands}

\begin{document}
\maketitle

\begin{abstract}
The dewetting of thin nanofilms is significantly impacted by thermal fluctuations, liquid-solid slip, and disjoining pressure, which can be described by lubrication equations augmented by appropriately scaled noise terms, known as stochastic lubrication equations. Here molecular dynamics simulations along with a newly proposed slip-generating method are adopted to study the instability of nanofilms with arbitrary slip. These simulations show that strong-slip dewetting is distinct from weak-slip dewetting by faster growth of perturbations and fewer droplets after dewetting, which can not be predicted by the existing stochastic lubrication equation. A new stochastic lubrication equation considering the strong slip boundary condition is thus derived using a long-wave approximation to the equations of fluctuating hydrodynamics. The linear stability analysis of this equation, i.e., surface spectrum, agrees well with molecular simulations. Interestingly, strong slip can break down the usual Stokes limits adopted in weak-slip dewetting and bring the inertia into effect. The evolution of the standard deviation of the film height $W^2(t)={\overline{h^2}-{\overline{h}}^2}$ at the initial stage of the strong-slip dewetting is found to be $W\sim t^{1/4}$ in contrast to $W\sim t^{1/8}$ for the weak-slip dewetting.
\end{abstract}
\section{Introduction}\label{sec1}
Nanometric thin liquid films deposited on substrates exist in a host of applications such as in lubricants\,\citep{jh1999}, coatings\,\citep{weinstein2004coating}, and microfluidics\,\citep{stone2004engineering}. The reliability of those applications depends heavily on our understanding of their stability mechanism, which is usually investigated in the context of thin-film flows\,\citep{or1997,craster2009dynamics}. Thin-film flows are characterized by the disparity of length scale in different dimensions, i.e., the ratio of film height $h$ to characteristic film length $\lambda$ is very small: $\chi= h/\lambda \ll1 $. This allows the adoption of a long-wave theory to derive lubrication equations from the full governing equations and boundary conditions, reducing the dimensionality and complexity of the problems\,\citep{or1997,craster2009dynamics}. 

Polymeric or metallic films on substrates with thicknesses below one hundred nanometers have been observed to undergo spontaneous rupture and dewetting\,\citep{xie1998spinodal,seemann2001dewetting,be2003,gonzalez2016inertial}. The dewetting mechanism in these films may be complicated due to the contamination of defects in the liquid. However, the primary dewetting mechanism for homogeneous liquid films is usually called spinodal dewetting. In this process, disjoining pressure, as a result of intermolecular forces between liquid and solid, leads to the instability of films. Basically, from the classical perspective, thermally excited capillary waves can be amplified by the disjoining
pressure, but in competition with the restoring force of surface tension, such that disturbances above a critical wavelength can grow and lead to film rupture\,\citep{vr1968}.

For interfacial flows at the nanoscale, thermal fluctuations can play an important role in the instability process\,\citep{mo2000,gr2006,zhang2019}. Thermal fluctuations can spontaneously generate thermal capillary waves (TCW) and roughness on the free surface of a liquid film at rest. The magnitude of thermal roughness is usually proportional to thermal length $\sqrt{k_BT/\gamma}$ ($k_B$, $T$, and $\gamma$ are the Boltzmann constant, temperature, and surface tension respectively)\,\citep{buff1965interfacial,ma2017}. Though the roughness is small and usually on the scale of nanometres, it becomes comparable to the size of films when the film thickness goes down to several nanometers. Note that micrometer roughness can also be obtained and thus observed optically in real space using ultra-low surface tension mixtures ($\gamma \sim 10^{-6}$ N/m)\,\citep{aa2004}. In the equilibrium state, the amplitude of TCW, known as the static spectrum, can be described by the renowned capillary wave theory\,\citep{buff1965interfacial,ma2017,hofling2015enhanced,hofling2020finite,hofling2024structure}. Recently, an extension of the capillary wave theory has been proposed utilizing a Langevin equation to describe the transient dynamics of non-equilibrium TCW and their approach to thermal equilibrium\,\citep{zhang2020thermal}. This advancement has led to the identification of a universality class governing the roughening behavior of film surfaces\,\citep{zhang2020thermal}.

The increasing importance of thermal fluctuations as the film height decreases may make the deterministic description of hydrodynamics at the nanoscale break down. For example, the breakup of liquid nanojets in molecular dynamics (MD) simulations\,\citep{mo2000,zh2019} and experiments\,\citep{hennequin2006drop} shows a double-cone rupture profile, in contrast to the long-thread profile predicted by the deterministic lubrication equation\,\citep{eggers1994drop}. \citet{mo2000} pioneered in showing that the deficiency of this deterministic lubrication equation for describing nanojet dynamics can be remedied by adding a noise term of appropriate strength to the equation, which leads to a stochastic lubrication equation for nanojets. \citet{eg2002} later shows that the evolution of the minimum neck radius is accelerated by thermal fluctuations, leading to $h_{min}\propto (t_0-t)^{0.418}$ ($t_0$ is the rupture time) in contrast with $h_{min}\propto (t_0-t)$ for the deterministic pinching. This noise-dominated breakup for nanojets has been observed in experiments using ultra-low surface tension mixtures\,\citep{hennequin2006drop}. 

For nanofilm rupture, \cite{gr2006} and \cite{da2005} independently derived the same stochastic lubrication equation for liquid films on \emph{no-slip} substrates. The numerical solution to this equation\,\citep{gr2006} can resolve the discrepancy in dewetting time between experimental results\,\citep{be2003} and the solution to the deterministic counterpart. Subsequently, the rupture of thin films with the effects of thermal fluctuations has been widely investigated by numerical solutions to the stochastic film equation\,\citep{gr2006,ne2015,di2016,du2019,sh2019}. These studies have consistently demonstrated that thermal fluctuations indeed accelerate the rupture process. The application of this stochastic film equation is not restricted to nanofilm dewetting. It has been extended to study, for example, nanodroplet spreading under an elastic sheet\,\citep{carlson2018fluctuation}, curvature-induced film drainage\,\citep{sh2019}, and mediated diffusion of particles confined in channels with a fluctuating wall\,\citep{marbach2018transport}. 

In addition to numerical solutions, linear stability analyses of the stochastic film equation have also been studied a lot, which allows us to obtain the evolution of the capillary spectra of surface waves\,\citep{me2005,fe2007,zh2019,zhang2019}. The analytical spectra show thermal fluctuations can massively amplify the growth of waves, shift the critical wavenumber to a larger value, and cause the dominant wavelength to evolve in time (in contrast to a constant value predicted by the deterministic lubrication equation). These interesting findings have been validated both in MD simulations\,\citep{zhang2019} and experiments\,\citep{fe2007}.

The original stochastic film equation mentioned above adopts the classical \emph{no-slip} boundary condition. As the flow scale reaches nanometers, surface effects like liquid-sold slip can have significant effects on flow behaviours (see the reviews by \citet{la2005,bo2010}). Obviously, nanofilm flows can be significantly impacted by slip as well, since the ratio of slip length $b$ to the film thickness $h$ can get close to unity or even much larger than unity\,\citep{baumchen2009slip}. In fact, in the deterministic setting, the introduction of slip to the deterministic film equation has been extensively studied for various phenomena, such as in coating a plate\,\citep{liao2013drastic}, droplet spreading\,\citep{savva2010two}, film rupture\,\citep{martinez2020effect}, falling films down a slippery plate\,\citep{ding2015falling}. However, only recently, the no-slip stochastic film equation is generalized to consider slip\,\citep{zhang2020}, which is non-trivial and requires the usage of the Green-Kubo-type expression\,\citep{bocquet1994hydrodynamic} that relates slip length to the random stress tensor at the wall. The derived slip equation is validated by the well-controlled molecular simulations\,\citep{zhang2020}.

However, the derived slip equation\,\citep{zhang2020} is limited to the case of \emph{weak slip} $b/h \approx 1$. In many cases, the slip length can be as large as micrometers so that $b/h\gg 1$, such as flow over graphene sheets \,\citep{falk2010molecular}, flow over engineered hydrophobic materials\,\citep{rothstein2010slip}, flow over substrates in presence of gas cavities and surface nanobubbles\,\citep{lohse2015surface}. For polymer liquids, increasing molecular weights can also increase the slip length up to micrometers\,\citep{baumchen2009reduced}. In fact, the dewetting of polymeric films on dodecyltrichlorosilane substrate (DTS), where slip length can be up to one micrometer, has been examined extensively in the deterministic framework\,\citep{fetzer2005new,munch2005lubrication,kargupta2004instability,
baumchen2014influence}. Notably, different levels of slip give rise to different deterministic lubrication models\,\citep{munch2005lubrication}. However, as we mentioned earlier, the effects of thermal fluctuations on nanofilms are significant, so a generalization of our current understanding of the instability of strong-slip films to consider thermal fluctuations is essential.

{So far, experimental studies on the effects of thermal fluctuations on thin film flows are limited due to the technical difficulties associated with the spatiotemporal scale. Though the experiments of the dewetting of polymer nanofilms on no-slip SiO\textsubscript{2}-coated silicon wafers have demonstrated the great effects of thermal fluctuations \citep{fe2007} on the growth of surface perturbations, dewetting of polymer films with strong slip and thermal fluctuations have not been considered in experiments. As such, molecular dynamics (MD) simulations are a natural and convenient tool to investigate the thin-film problem at the nanoscale as thermal fluctuations are inherent in MD simulations}. 

In this work, molecular dynamics simulations are employed to simulate the rupture of nanofilms on substrates with a strong slip, in comparison with the small-slip rupture. A new simulation strategy is proposed to generate strong slip in molecular simulations since the classic molecular simulations are limited to weak slip. We obtain the evolution of film surface spectra, rupture time, and number of droplets after film rupture from molecular simulations. A new stochastic lubrication equation considering the strong slip is derived from fluctuating hydrodynamics. A linear stability analysis of this stochastic lubrication equation leads to the analytical spectra which are validated against the MD results to establish the applicability of the new theory to predict future experiments.

This paper is organized as follows. In \S\,\ref{sec2}, the stochastic lubrication equation for the strong-slip dewetting is derived from the equations of fluctuating hydrodynamics using a long-wave approximation. In \S\,\ref{sec3}, a linear stability analysis of the newly derived stochastic equation is performed to obtain the surface spectrum. In \S\,\ref{sec4}, molecular simulations of the rupture of nanofilms with the method to generate strong slip are presented. \S\,\ref{sec5} compares the new model with molecular simulation results, and discusses new findings. In \S\ref{sec6}, we summarise the main contributions of this work and outline future directions of research.
\section{Stochastic lubrication equation for films with strong slip}\label{sec2}
\subsection{Governing equations and boundary conditions}
\begin{figure}
\includegraphics[width=\linewidth]{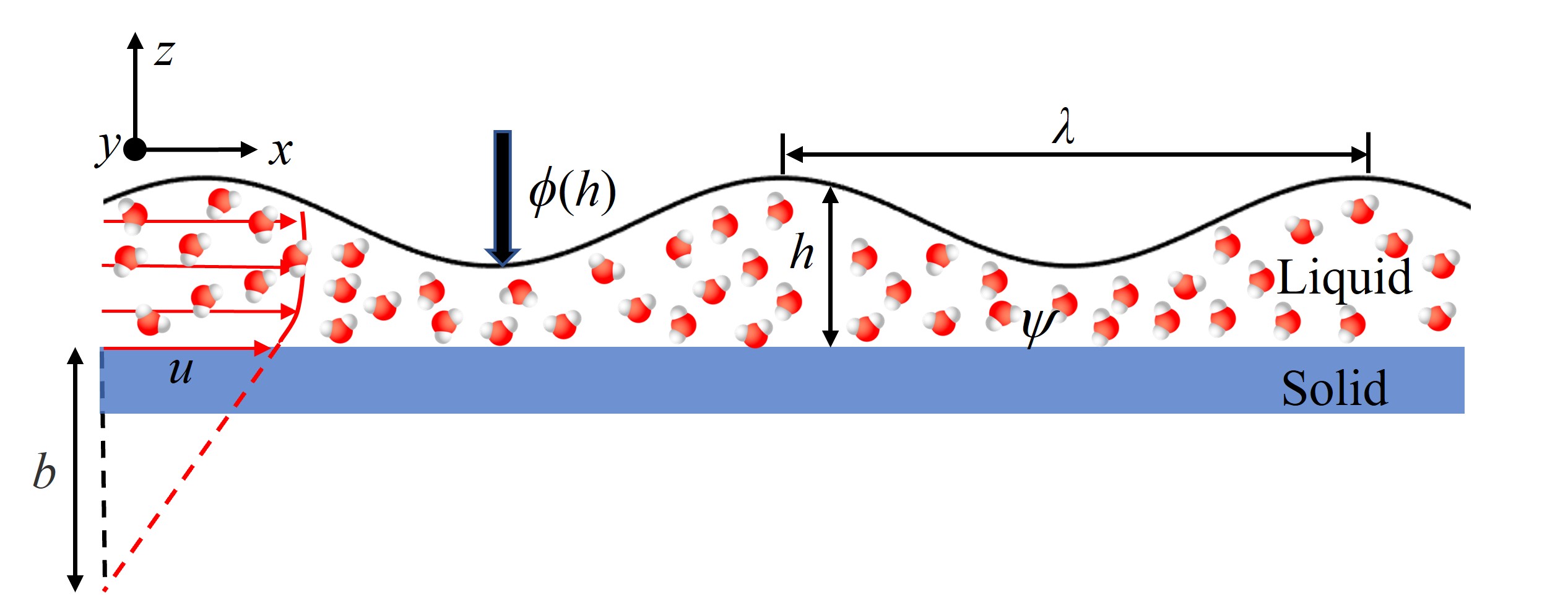}
\caption{Sketch of a (quasi-two dimensional) molecularly thin liquid film on a slippery substrate. Here $h(x,t)$ is the film thickness, $\lambda$ is the characteristic length, $u$ is the horizontal velocity, $\phi$ is the disjoining pressure, and $b$ is the slip length. The film has a small depth $L_y$ into the page.}
\label{fig1}
\end{figure}
As shown in figure \ref{fig1}, a molecularly thin liquid film is deposited on a solid surface and destabilized by both disjoining pressure $\phi$ and thermal fluctuations $\psi$. The film is quasi-two dimensional (2D) confined by its size ($L_x$, $L_y$, $h$) where $L_x\gg L_y$ and $L_x\gg h$. Without slip, the stochastic thin film equation is derived in detail by \citet{gr2006} using a long-wave approximation ($\chi=h_0/\lambda \ll 1$) to fluctuating hydrodynamics (FH)\,\citep{la1959}. The no-slip stochastic equation has been extended to consider weak slip $b\sim \mathcal{O}(h)$\,\citep{zhang2020}. Here we present the derivation of a new equation considering $b\gg h$. 

The governing equations for this problem are given by equations of FH, where thermal fluctuations are modelled by an additional random stress tensor. The (incompressible) continuum equation and momentum equations are 
\begin{equation}\label{eq_nonconti}
{\partial_x}u+{\partial_z}w=0,
\end{equation}
\begin{equation} \label{eq_nonmom1}
   \rho \left({\partial_t}u+u{\partial_xu}+w{\partial_zu} \right)=-{\partial_x p}+\mu \left( {\partial_{xx}u}+{\partial_{{zz}}u} \right) +{\partial_x}{{\psi }_{xx}}+{\partial_z}{{\psi }_{zx}},
\end{equation}
\begin{equation} \label{eq_nonmom2}
 \rho \left( {\partial_t}w+u {\partial_x w}+w{\partial_zw} \right)=-{\partial_z p}+\mu \left( {\partial_{{xx}}w}+{\partial_{{zz}}}w \right) +{\partial_x}{{\psi }_{xz}}+{\partial_z}{{\psi }_{zz}}.
\end{equation}
Here $u$ and $w$ are the $x$ and $z$ components of velocity, and $\psi$ is a 2D random stress tensor with zero mean and covariance given by
\begin{equation}
{{\left\langle \psi  \right.}_{ij}}(\bm{x},t)\left. {{\psi }_{lm}}(\bm{x}',t') \right\rangle= \frac{2\mu {{k}_{B}}\theta}{L_y}\left( {{\delta }_{il}}{{\delta }_{jm}}+{{\delta }_{im}}{{\delta }_{jl}} \right) 
\delta \left({x}-{x}' \right)\delta \left({z}-{z}'\right)\delta(t-t').
\end{equation}
Here $\theta$ is the temperature. The factor $1/L_y$ appears because the films are quasi-2D ($L_y \ll L_x$), allowing all variables of interest to be averaged over the $y$ direction\,\citep{zhang2020}.

For boundary conditions, at $z=h$, we have the dynamic condition and kinematic condition:
\begin{subequations}
\begin{equation}
(\bm{\sigma}+\bm{\psi})\cdot\mathbf{n}=-[\gamma\nabla\cdot\mathbf{n}+\phi(h)]\mathbf{n}, \quad 
{\partial_t h}+u{\partial_x h}=w \quad \mbox{at\ }\quad z=h. \tag{2.5a,b}
\end{equation}
\end{subequations}
Here $\bm{\sigma}=-p{{\delta }_{ij}}+\mu \left(\partial_{{x}_{j}}{u}_{i}+\partial_{{x}_{i}}{{u}_{j}} \right)$ is the hydrodynamic stress tensor {(here $i=(x,z)$ and $j=(x,z)$)}, $\gamma$ is the surface tension, $\phi(h)$ is the disjoining pressure, and $\mathbf{n}$ is the outer normal vector at the surface $\mathbf{n}={\left( -\partial_x h,1 \right)}/{\sqrt{1+{{\left(\partial_x h \right)}^{2}}}}$.

At $z=0$, the impermeable condition and Navier's slip boundary condition are separately given by
\begin{subequations}
\begin{align} \label{eq_nonsb}
w=0, \quad 
u=b\frac{\partial u}{\partial z}, \quad \mbox{at\ }\quad z=0. 
\tag{2.6a,b}
\end{align}
\end{subequations}
where $b$ is the slip length. The covariance of the random shear stress at the wall is given by\,\citep{bocquet1994hydrodynamic,zhang2020}: 
\begin{equation}
\left\langle\psi_{zx}{{|}_{z=0}}(x,t)\psi'_{zx}{{|}_{z=0}}(x',t')\right\rangle=\frac{2\mu {{k}_{B}}\theta}{b L_y}\delta (x-x')\delta (t-t').
\end{equation}
{Navier's slip condition is chosen because it has been extensively validated by both experiments and MD simulations (see the reviews by \citet{la2005,bo2010}). However, there are many other forms of slip boundary conditions, as discussed by \citet{sibley2015comparison}.}
\subsection{A long wave approximation}
To derive a lubrication equation, \eqref{eq_nonconti}-\eqref{eq_nonsb} have to be scaled based on the dominant mechanism of momentum balance, which varies with the level of slip length\,\citep{munch2005lubrication}. Classically, for the weak-slip case, the momentum balance happens in the horizontal direction $\partial_x p \sim\mu \partial_{zz}u$, which means $ph_0/(\mu u_0) \sim 1/\chi$, where $u_0$ is the characteristic velocity. At the free surface, surface tension has to be balanced with pressure $p=-\gamma \partial_{xx} h $, which means $\gamma/(\mu u_0)\sim 1/\chi^3$. At the solid surface, the order of slip length is $b\sim h_0$. Therefore, the pressure term, surface tension term, and slip length will be scaled to $P=\chi ph_0/(\mu u_0)$, $\Gamma=\chi^3 \gamma/(\mu u_0)$, and $B=b/h_0$.

However, for the strong-slip (including free slip) flow, the velocity profile essentially becomes uniform (plug flow) instead of being parabolic so that the momentum balance happens in the vertical direction $\partial_z p \sim\mu \partial_{zz}w$\,\citep{munch2005lubrication}, which leads to the scaling $ph_0/(\mu u_0) \sim \chi$ and $\gamma/(\mu u_0)\sim 1/\chi$. These scalings need the slip length to be strong and $b\sim h_0\chi^{-2}$\,\citep{munch2005lubrication}.

As for the scaling of the random stress tensor, it may be scaled the same as the scale of their deterministic counterparts\,\citep{gr2006}. For example,  $\psi_{xx}\sim \mu \partial_x u \sim \mu u_0/\lambda$ and $\psi_{zz}\sim \mu \partial_z w \sim \mu u_0/\lambda$. Note in this way, two scaling exists for $\psi_{zx}$: $\psi_{zx}\sim \mu \partial_x w \sim \mu \chi u_0/\lambda$ and $\psi_{zx}\sim \mu \partial_z u \sim \mu u_0/(\chi\lambda)$. The former one is used to keep $\psi_{zx }$ at a lower order.  In summary, the following scalings are used:
\begin{gather}
  X=\frac{x}{\lambda },Z=\frac{z}{{{h}_{0}}},H=\frac{h}{{{h}_{0}}},B=\frac{b}{{{h}_{0}}\chi^{-2}},U=\frac{u}{{{u}_{0}}},W=\frac{w}{\chi {{u}_{0}}},T=\frac{{{u}_{0}}t}{\lambda },\Gamma =\frac{\chi \gamma }{\mu {{u}_{0}}} \nonumber, \\ 
  \left( P,\Phi  \right)=\frac{{{h}_{0}}}{\chi {{u}_{0}}\mu }\left( p,\phi  \right),\left( {{\Psi }_{xx}},{{\Psi }_{zz}} \right)=\frac{\lambda }{{{u}_{0}}\mu }\left( {{\psi }_{xx}},{{\psi }_{zz}} \right),\left( {{\Psi }_{xz}},{{\Psi }_{zx}} \right)=\frac{{\lambda}}{{{\chi }}{{u}_{0}}\mu }\left( {{\psi }_{xz}},{{\psi }_{zx}} \right) . \label{eq_scaling}
\end{gather}

Adopting these scalings \eqref{eq_scaling}, the rescaled and dimensionless continuum and momentum equations are
\begin{equation}
{\partial_X U}+{\partial_Z W}=0,
\end{equation}
\begin{equation}
\chi \operatorname{Re}\left({\partial_T U}+U{\partial_X U}+W{\partial_Z U} \right)=-{{\chi }^{2}}{\partial_X P}+ {{\chi }^{2}}\partial_{{XX}}U+{\partial_{{ZZ}}}U+{{\chi }^{2}}{\partial_X {{\Psi }_{xx}}}+{{\chi }^{2}}{\partial_Z {{\Psi }_{zx}}},
\end{equation}
\begin{equation}
\chi \operatorname{Re}\left({\partial_T W}+U{\partial_X W}+W{\partial_{Z} W} \right)=-{\partial_Z P}+{{\chi }^{2}}{\partial_{{XX}}W}+{\partial_{{ZZ}}W} + {{\chi }^{2}}{\partial_X {{\Psi }_{xz}}}+{\partial_Z {{\Psi }_{zz}}}.
\end{equation}
Here the Reynolds number is $\text{Re}={\rho u_0 h_0 }/{\mu}$. As the characteristic velocity is $u_0\sim\chi{\gamma}/{\mu}$, the $\text{Re}$ can be also written as $\text{Re}=\chi{\rho \gamma h_0 }/{\mu^2}=\chi \text {La}$, where $La={\rho \gamma h_0 }/{\mu^2}$ is the Laplace number. The $\text{La}$ is assumed to be of order one.

In terms of the rescaled boundary conditions,
\begin{subequations}
\begin{gather}
    -P+\frac{1}{1+{{\chi }^{2}}{{\left( {{\partial }_{X}}H \right)}^{2}}}\left\{ 2\left[ 1-{{\chi }^{2}}{{\left( {{\partial }_{X}}H \right)}^{2}} \right]{{\partial }_{Z}}  W-2{{\partial }_{X}}H\left( {{\partial }_{Z}}U+{{\chi }^{2}}{{\partial }_{X}}W \right)+{{\chi }^{2}}{{\left( {{\partial }_{X}}H \right)}^{2}}{{\Psi }_{xx}}-  \right. \nonumber\\  
\left.{{\chi }^{2}}{{\partial }_{X}}H\left( {{\Psi }_{xz}}+{{\Psi }_{zx}} \right)+{{\Psi }_{zz}} \right\} =\Gamma \frac{{{\partial }_{XX}}H}{{{\left[ 1+{{\chi }^{2}}{{\left( {{\partial }_{X}}H \right)}^{2}} \right]}^{3/2}}}+\Phi  \quad \mbox{at\ }\quad z=H, 
\end{gather}
\begin{gather}
  2{{\chi }^{2}}{\partial_X H}\left[ \left( {\partial_Z W}-{\partial_X U} \right)+\frac{1}{2}\left( {{\Psi }_{zz}}-{{\Psi }_{xx}} \right) \right]+\left[ 1-{{\chi }^{2}}{{\left({\partial_X H} \right)}^{2}} \right]\left( {\partial_Z U}+{{\chi }^{2}}{\partial_X W}+{{\chi }^{2}}{{\Psi }_{zx}} \right)=0\nonumber \\
 \quad \mbox{at\ }\quad z=H ,
\end{gather}
\begin{equation}
  {\partial_T H}+U{\partial_X H}=W \quad \mbox{at\ }\quad z=H.
\end{equation}
\begin{equation}
W=0\quad \mbox{at\ }\quad z=0,
\end{equation}
\begin{equation}
U=\frac{B}{{{\chi }^{-2 }}}{\partial_Z U}\quad \mbox{at\ }\quad z=0.
\end{equation}
\end{subequations}

The rescaled equations can be approximately solved by the perturbation expansion of $U,W,P,\Psi,H$:
\begin{equation} \label{eq_exp}
\left( U,W,P,\Psi,H  \right)=\left( {{U}_{0}},{{W}_{0}},{{P}_{0}},{{\Psi }_{0}},H_0 \right)+{{\chi }^{2}}\left( {{U}_{1}},{{W}_{1}},{{P}_{1}},{{\Psi }_{1}},H_1 \right)+...
\end{equation}

Then, at leading orders of governing equations, one can get 
\begin{equation}
   {\partial_{{ZZ}}U_0}=0 ,
\end{equation}
\begin{equation}
  {\partial_{{ZZ}}W_0}-{\partial_Z P_0}+{\partial_Z {{\Psi }_{zz0}}}=0, 
\end{equation}
\begin{equation}
 {\partial_X {{U}_{0}}}+{\partial_Z {{W}_{0}}}=0,
\end{equation}
For boundary conditions, their leading-order forms are
\begin{subequations}
\begin{equation}
   -{{P}_{0}}+2\left( {\partial_Z {{W}_{0}}}-{\partial_X H_0}{\partial_Z {{U}_{0}}}\right)+{{\Psi }_{zz0}}=\Gamma {{\partial_{XX} }}H_0+\Phi \quad \mbox{at\ }\quad z=H,
 \end{equation}
\begin{equation}
 {\partial_Z {{U}_{0}}}=0 \quad \mbox{at\ }\quad z=H,
 \end{equation} 
\begin{equation}
  {{W}_{0}}={\partial_T {{H}_{0}}}+{{U}_{0}}{\partial_X {{H}_{0}}}\quad \mbox{at\ }\quad z=H,
 \end{equation}
\begin{equation}
W_0=0 \quad \mbox{at\ }\quad z=0,
\end{equation}
\begin{equation}
\partial_Z U_0=0 \quad \mbox{at\ }\quad z=0.
\end{equation}
\end{subequations}

Using leading order equations, we find
\begin{align}
 {{U}_{0}} & \equiv {{U}_{0}}(X,T), \\ 
 {{W}_{0}} &=-{\partial_X {{U}_{0}}}Z, \\ 
  {{P}_{0}}&=-2{\partial_X U_0}-\Gamma{\partial_{XX}H_0}-\Phi +{{\Psi }_{zz0}} ,
\end{align}
which leads to the first equation of the desired stochastic lubrication equation:
\begin{equation}\label{nondimsle1}
{\partial_T {{H}_{0}}}+{\partial_X \left( {{H}_{0}}{{U}_{0}} \right)}=0.
\end{equation}

In the next order, the governing equation that will be used is,
\begin{equation}\label{2ndge}
\mathrm{La}\left({\partial_T {{U}_{0}}}+{{U}_{0}}{\partial_X {{U}_{0}}}\right)={{{\partial}_{XX}}{{U}_{0}}}-{\partial_X {{P}_{0}}}\ + {{\partial_{ZZ} }}{{U}_{1}}+{\partial_X {{\Psi }_{xx0}}}+{\partial_Z {{\Psi }_{zx0}}}
\end{equation}
and the boundary condition that will be needed is 
\begin{equation}\label{eq2.22}
   -4{\partial_X H_0}{\partial_X U_0}+{\partial_X H_0}\left( {{\Psi }_{zz0}}-{{\Psi }_{xx0}} \right)+ {\partial_Z {{U}_{1}}}-H_0{{{\partial_{XX} }}U_0}+{{\Psi }_{zx0}}=0, \quad \text{at}\quad z=H,
\end{equation}
\begin{equation}\label{eq2.23}
 {{U}_{0}}=B{\partial_Z {{U}_{1}}},\quad \text{at}\quad z=0.
\end{equation}

Integrating \eqref{2ndge} from $0$ to $H_0$ and using boundary conditions \eqref{eq2.22} and \eqref{eq2.23}   leads to
\begin{align}\label{nondimsle2}
  &H_0\left[ \operatorname{La}\left({\partial_T {{U}_{0}}}+{{U}_{0}}{\partial_X {{U}_{0}}}\right) \right]  =H_0\left[ 3{{{\partial_{XX} }}{{U}_{0}}}+{\partial_X }\left( \Gamma {{{\partial }_{XX}}H_0}+\Phi  \right) \right]\nonumber \\
&+\left({\partial_Z {{U}_{1}}}+{{\Psi }_{zx0}} \right){{|}_{Z=H_0}}-\left({\partial_Z {{U}_{1}}}+{{\Psi }_{zx0}} \right){{|}_{Z=0}}+\int_{0}^{H_0}{{\partial_X }\left( {{\Psi }_{xx0}}-{{\Psi }_{zz0}} \right)dZ} \nonumber\\ 
 & =H_0{\partial_X }\left(\Gamma{\partial_{XX}H_0}+\Phi  \right)+4{\partial_X }\left( H_0{\partial_X {{U}_{0}}} \right)-\frac{{{U}_{0}}}{B}+{\partial_X}\int_{0}^{H_0}{\left( {{\Psi }_{xx0}}-{{\Psi }_{zz0}} \right)dZ}-{{\Psi }_{zx0}}{{|}_{Z=0}} .
\end{align}
The Leibniz integral rule is used here. The covariance of $\Psi _{zx0}{|}_{Z=0}$ is given by the Green-Kubo expression for slip length $\left\langle\Psi_{zx}{{|}_{Z=0}}(X,T)\Psi'_{zx}{{|}_{Z=0}}(X',T')\right\rangle=\frac{2 {{k}_{B}}\theta}{\chi^2 \mu u_0 b L_y}\delta (X-X')\delta (T-T')$\,\citep{zhang2020}.
 In terms of the integral of the white noise in \eqref{nondimsle2}, as ${\Psi }_{xx0}$ is uncorrelated with ${\Psi }_{zz0}$, they can be combined as $\sqrt2{\Psi }_{xx0}$. Now using the theorem provided in the appendix of \citet{zhang2020}, the integral of the white noise is
\begin{align}
  {\partial_X }\int_{0}^{H_0}{\left( {{\Psi }_{xx0}}-{{\Psi }_{zz0}} \right)dz}  &=\sqrt{2}{\partial_X }\int_{0}^{H_0}{ {{\Psi }_{xx0}} \, dz} \nonumber\\ 
 & =\sqrt{2}{\partial_X }\left( \sqrt{H_0}\Re  \right) ,
\end{align}
where $\left\langle {{\Re(X,T)}}{{\Re(X',T') }}' \right\rangle =\frac{4 {{k}_{B}}\theta}{\mu  u_0 h_0 L_y}\delta \left( X-X' \right)\delta \left( T-T' \right) $.

Putting \eqref{nondimsle1} and \eqref{nondimsle2}  together and returning to their dimensional forms, we arrive at the stochastic lubrication equation for films with strong slip (referred to as S-S model hereafter) 
 \begin{subequations} \label{sle12}
\begin{gather}
{\partial_t h}+{\partial_x \left( hu \right)}=0,  \\ 
   \rho \left({\partial_t u}+u{\partial_x u}\right)={\partial_x}\left( -\gamma {{{\partial_{xx} }}h}+\phi  \right)+\frac{4\mu }{h}{\partial_x }\left(h{\partial_x u}\right)-\frac{\mu }{h}\frac{u}{b}+\frac{1}{h}\left[ 2 {\partial_x }\left( \sqrt{h}{{\xi }_{1}} \right)-\frac{\sqrt b}{b}{{\xi }_{2}} \right]
\end{gather}
 \end{subequations}
where the covariance of the noise term is $\left\langle {{\xi_i(x,t) }}{{\xi_j (x',t')}} \right\rangle =\frac{2\mu {{k}_{B}}\theta}{L_y}\delta_{ij}\delta \left( x-x' \right)\delta \left( t-t' \right) $. 

The above derived stochastic lubrication equation validates for the strong slip length on the order of $b\sim h_0\chi^{-2}$. In terms of the weak slip $b\sim h_0$, a similar long-wave approximation to FH equations (see \citet{zhang2020} for details) can result in the weak-slip stochastic lubrication equation (referred to as W-S model hereafter)
\begin{equation}
{\partial_t h}=\frac{1}{\mu }{\partial_x }\left[\left(\frac{1}{3}{{h}^{3}}+b{{h}^{2}}\right){\partial_x}\left(-\gamma{\partial_{xx} h}+\phi\right)+\sqrt{\frac{1}{3}{{h}^{3}}+b{{h}^{2}}}\xi_3  \right],
\end{equation}
where the noise $\xi_3$ has zero mean and covariance $\left\langle \xi_3(x,t)\xi_3(x',t')\right\rangle=\frac{2\mu k_B \theta}{L_y}\delta(x-x')\delta(t-t')$. As the W-S model works for the no-slip case ($b=0$), the S-S model validates for the free-slip case ($b=\infty$) where the terms containing $b$ in \eqref{sle12} vanish. Interestingly, this free-slip model can be applied to study free films such as foam films\,\citep{erneux1993nonlinear,vaynblat2001rupture}. {Note that the free-slip model has not been reported before.}

\section{Surface spectrum for films with strong slip}\label{sec3}
With the newly developed S-S model \eqref{sle12}, the spectrum of surface waves in linear stages is derived here. We firstly rewrite the noise $\xi$ in terms of the white noise $N$ with unit variance $\left\langle {{N_i}}{{N_j }}' \right\rangle =\delta_{ij}\delta \left( x-x' \right)\delta \left( t-t' \right)$, and linearise \eqref{sle12} with $h=h_0+\tilde{h}$, $u=0+\tilde{u}$, and $N=0+\tilde{N}$:
\begin{equation} \label{eq_lsle12}
\frac{{{\partial }^{2}}\tilde{h}}{\partial {{t}^{2}}}=\frac{\mu }{\rho }\frac{\partial }{\partial t}\left[ 4\frac{{{\partial }^{2}}\tilde{h}}{\partial {{x}^{2}}}-\frac{{\tilde{h}}}{{{h}_{0}}b} \right]-\frac{{{h}_{0}}}{\rho }\left[ \frac{\partial \phi }{\partial h}\frac{{{\partial }^{2}}\tilde{h}}{\partial {{x}^{2}}}+\gamma \frac{{{\partial }^{4}}\tilde{h}}{\partial {{x}^{4}}} \right]-f_1\frac{{{\partial }^{2}}{\tilde{N}_{1}}}{\partial {{x}^{2}}}+f_2\frac{\partial {\tilde{N}_{2}}}{\partial x},
\end{equation}
where the factor ${{f}_{1}}=\sqrt{8\mu {{k}_{B}}\theta{{h}_{0}L_y}}/{\left(\rho L_y\right) }$ and ${{f}_{2}}=\sqrt{{2\mu {{k}_{B}}\theta}{b}L_y}/{\left(\rho b L_y\right) }$. {Note that the second derivative of $\tilde{h}$ with respect to $t$ comes from putting the linearized \eqref{sle12}(a) into the linearized \eqref{sle12}(b) to eliminate the variable $\tilde{u}$.} 
Then a Fourier transform of \eqref{eq_lsle12} is performed using $\widehat{h}(q,t)=\int_{-\infty }^{\infty }{\tilde{h}(x,t){{e}^{-}}^{iqx}d}x$ and $\widehat{N}(q,t)=\int_{-\infty }^{\infty }{\tilde{N}(x,t){{e}^{-}}^{iqx}d}x$ to get 
\begin{equation}\label{eqlsan1n2}
\frac{{{\partial }^{2}}\widehat{h}}{\partial {{t}^{2}}}=-C\frac{\partial \widehat{h}}{\partial t}-D\widehat{h}+f_1{{q}^{2}}{{\hat{N}}_{1}}+f_2 qi{{\hat{N}}_{2}}.
\end{equation}
Here $C=\frac{\mu }{\rho }\left( 4{{q}^{2}}+\frac{1}{{{h}_{0}}b} \right)$, $D=\frac{{{h}_{0}}}{\rho }\left[ \frac{\partial \phi }{\partial h}{{q}^{2}}+\gamma {{q}^{4}} \right]$.
The solution of \eqref{eqlsan1n2} can be represented as the linear superposition of two contributions\,\citep{zh2019,zhang2019}
\begin{equation}
\hat h={\hat h}_{\mathrm{det}}+{\hat h}_{\mathrm{sto}},
\end{equation}
where ${{\hat{h}}_{\mathrm{det}}}$ is the solution to the deterministic part of (\ref{eqlsan1n2}) and ${{\hat{\ h}}_{\mathrm{sto}}}$ is the contribution purely caused by thermal fluctuations. To find ${{\hat{h}}_{\mathrm{det}}}$, the deterministic equation
\begin{equation}
\frac{{{\partial }^{2}}\widehat{h}}{\partial {{t}^{2}}}+C\frac{\partial \widehat{h}}{\partial t}+D\widehat{h}=0,
\end{equation}
is solved by Laplace transform. Using $g(q,s)=\int_{0}^{\infty }{\widehat{h}(q,t){{e}^{-}}^{its}d}t$ and assuming $\frac{\partial \widehat{h}}{\partial t}{{|}_{t=0}}$, one can get
\begin{equation}
g=\frac{s+C}{{{s}^{2}}+Cs+D}\widehat{h}(q,0),
\end{equation}
whose inverse Laplace transform is
\begin{equation} \label{eqssdet}
{{\widehat{h}}_{\det }(q,t)}=\widehat{h}(q,0)\left[ \frac{{{e}^{{{\omega }_{1}}t}}+{{e}^{{{\omega }_{2}}t}}}{2}+\frac{C}{2\sqrt{{{C}^{2}}-4D}}\left( {{e}^{{{\omega }_{1}}t}}-{{e}^{{{\omega }_{2}}t}} \right) \right] 
\end{equation}
Here $\omega_{i=1,2}=\frac{-C\pm \sqrt{C^2-4D}}{2}$ is the solution to $s^2+Cs+D=0$. Notably, $\omega_1$ is the dispersion relation (growth rate) of the deterministic lubrication equation, namely, \eqref{sle12} without the noise terms. 

To obtain $\hat h_\mathrm{sto}$, one has to determine the impulse response of the linear system.  Using the Laplace transform of $\frac{{{\partial }^{2}}\widehat{h}}{\partial {{t}^{2}}}+C\frac{\partial \widehat{h}}{\partial t}+D\widehat{h}=\delta$, and assuming $\hat{h}(q,0)=0$, it is found
  \begin{equation}\label{eq_im_re}
g=\frac{1}{{{s}^{2}}+Cs+D}.
\end{equation}
The impulse response is thus the inverse Laplace transform of \eqref{eq_im_re}
\begin{equation}
F(q,t)={{\widehat{h}}}=\frac{{{e}^{{{\omega }_{1}}t}}-{{e}^{{{\omega }_{2}}t}}}{\sqrt{{{C}^{2}}-4D}}.
\end{equation}
Now with thermal fluctuations $f_1{{q}^{2}}\widehat{N}_1$ and $f_2 qi\widehat{N}_2$ as input, we find
\begin{equation}
{{\widehat{h}}_\mathrm{sto}}=f_1 q^2{{{}_{_{}}}\int_{0}^{t}{\widehat{N}_1\left( q,t-\tau  \right)}F(q,\tau )d\tau } +f_2 q i{{{}_{_{}}}\int_{0}^{t}{\widehat{N}_2\left( q,t-\tau  \right)}F(q,\tau )d\tau }\label{eqsssto}.
\end{equation}
As $\hat{h}$ is both a random and complex variable, the root
mean square (RMS) of its norm is sought, namely, surface spectrum,
\begin{equation}
{{\left| \widehat{ h}(q,t) \right|}_{\mathrm{rms}}}=\sqrt{\overline{{{\left| {{\widehat{ h}}_{\mathrm{det}}}+{{\widehat{ h}}_{\mathrm{sto}}} \right|}^{2}}}}  =\sqrt{{{\left| {{\widehat{ h}}_{\mathrm{det}}} \right|}^{2}}+\overline{{{\left| {{\widehat{ h}}_{\mathrm{sto}}} \right|}^{2}}}},
\end{equation}
where from \eqref{eqssdet}
\begin{equation}
{\left| {{\widehat{ h}}_{\mathrm{det}}} \right|}^{2}=\overline{{{\left| \widehat{h}(q,0) \right|}^{2}}}{{\left[ \frac{{{e}^{{{\omega }_{1}}t}}+{{e}^{{{\omega }_{2}}t}}}{2}+\frac{C}{2\sqrt{{{C}^{2}}-4D}}\left( {{e}^{{{\omega }_{1}}t}}-{{e}^{{{\omega }_{2}}t}} \right) \right]}^{2}} ,
\end{equation}
and from \eqref{eqsssto}
\begin{align}
   \overline{{{\left| {{\widehat{\ h}}_\mathrm{sto}} \right|}^{2}}}&={{{\left| {f_1 q^2}\int_{0}^{t}{\widehat{N}_1\left( q,t-\tau  \right)}F(q,\tau )d\tau  \right|}^{2}}} +{{{\left| {f_2 q i}\int_{0}^{t}{\widehat{N}_1\left( q,t-\tau  \right)}F(q,\tau )d\tau  \right|}^{2}}} \nonumber\\ 
 & ={{{\left| {f_1 q^2} \right|}^{2}}\int_{0}^{t}{{{\left| \widehat{N}_1\left( q,t-\tau  \right) \right|}^{2}}}F{{(q,\tau )}^{2}}d\tau }+{{{\left| {f_2 q i} \right|}^{2}}\int_{0}^{t}{{{\left| \widehat{N}_1\left( q,t-\tau  \right) \right|}^{2}}}F{{(q,\tau )}^{2}}d\tau } \nonumber\\ 
 & =\left( {{\left| {f_1 q^2} \right|}^{2}}+{{\left| {f_q q i} \right|}^{2}} \right){{L}_{x}}\int_{0}^{t}{F{{(q,\tau )}^{2}}}d\tau  \nonumber\\ 
 & =\frac{2\mu {{k}_{B}}T{{L}_{x}}{{q}^{2}}}{{{\rho }^{2}}{{L}_{y}}\left( {{C}^{2}}-4D \right)}\left( 4{{h}_{0}}{{q}^{2}}+\frac{1}{b} \right)\left[ \frac{{{e}^{2{{\omega }_{1}}t}}-1}{2{{\omega }_{1}}}+\frac{{{e}^{2{{\omega }_{2}}t}}-1}{2{{\omega }_{2}}}+\frac{2\left( {{e}^{-Ct}}-1 \right)}{C} \right].
\end{align}
Here we have used $\overline{{{\left| \widehat{N}\left( q,t \right) \right|}^{2}}}={{L}_{x}}$, due to the finite length of the discrete Fourier transform used in MD simulations. Thus, we derive the spectrum of surface waves of a bounded film with strong slip as,
\begin{align}\label{gsss}
 S\left(q,t\right)
 & =\left\lbrace\overline{{{\left| \widehat{h}(q,0) \right|}^{2}}}{{\left[ \frac{{{e}^{{{\omega }_{1}}t}}+{{e}^{{{\omega }_{2}}t}}}{2}+\frac{C}{2\sqrt{{{C}^{2}}-4D}}\left( {{e}^{{{\omega }_{1}}t}}-{{e}^{{{\omega }_{2}}t}} \right) \right]}^{2}} \nonumber\right.\\ 
 & \left.+\frac{2\mu {{k}_{B}}T{{L}_{x}}{{q}^{2}}}{{{\rho }^{2}}{{L}_{y}}\left( {{C}^{2}}-4D \right)}\left( 4{{h}_{0}}{{q}^{2}}+\frac{1}{b} \right)\left[ \frac{{{e}^{2{{\omega }_{1}}t}}-1}{2{{\omega }_{1}}}+\frac{{{e}^{2{{\omega }_{2}}t}}-1}{2{{\omega }_{2}}}+\frac{2\left( {{e}^{-Ct}}-1 \right)}{C} \right]\right\rbrace^{1/2}
\end{align}
\eqref{gsss} and \eqref{sle12} are main contributions of this work.
For the W-S model, its analytical spectrum is\,\citep{zhang2020}
\begin{align}\label{wsss}
  & S\left(q,t\right)=\sqrt{\overline{{{\left| \widehat{h}(q,0) \right|}^{2}}}{{e}^{2{{\omega }_{3}}(q)t}}+\frac{{{L}_{x}}}{{{L}_{y}}}\frac{{{k}_{B}}T}{\gamma {{q}^{2}}+\left.\left( d\phi /dh \right)\right.|_{h_0}}\left[1-{{e}^{2{{\omega }_{3}}(q)t}}\right]},
\end{align}
where the dispersion relation is
\begin{equation}
{{\omega }_{3}}=-\frac{h_0^3+bh_0^2}{3\mu }\left( \gamma {{q}^{4}}+\left.\frac{d \phi }{d h}\right.| _{{{h}_{0}}}{{q}^{2}}\right).
\end{equation} 
\section{Molecular simulations and a new slip-generating method} \label{sec4}
\begin{figure}
\includegraphics[width=\linewidth]{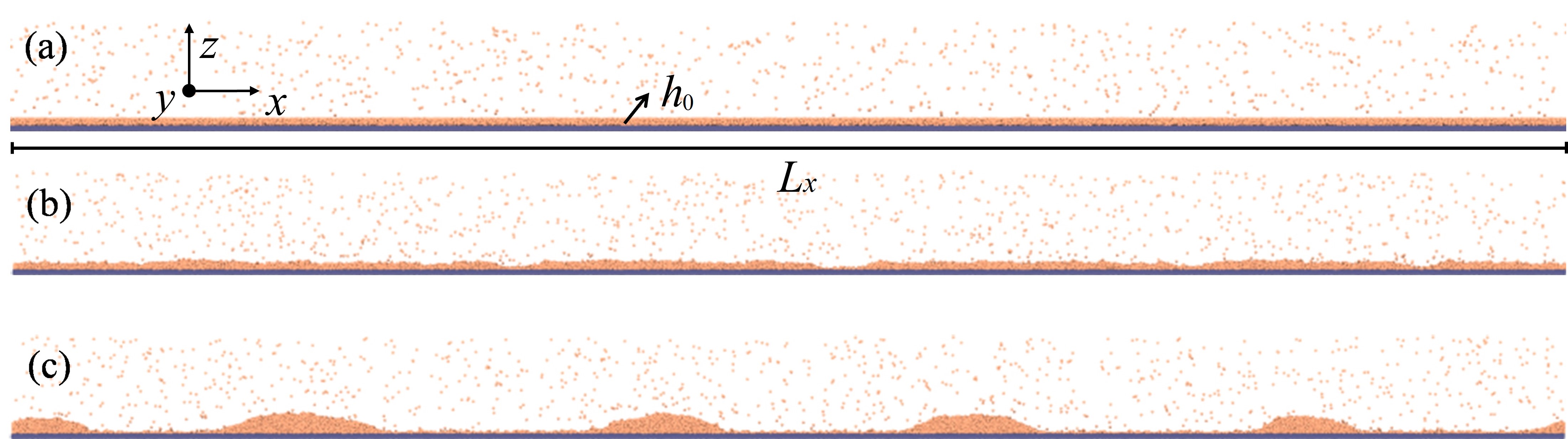}
\caption{Rupture of thin liquid films with strong slip in molecular simulations. (a) initial setting of a liquid film with a flat free surface. The film has a small depth $L_y$ into the page. (b) growth of perturbations and film rupture. (c) droplets formation after that the film ruptures. Note that the wall colored in blue only denotes the position of the solid boundary as the method of a virtual wall is used in simulations.}
\label{fig2}
\end{figure}

Molecular dynamics (MD) simulations are performed to simulate the instability of nanofilms on the strong-slip solid and the weak-slip solid. The open-source MD code LAMMPS\,\citep{pl1995} is adopted. As shown in figure \ref{fig2}(a), the liquid film is composed of liquid Argon (represented in orange) and it is simulated with the standard Lennard-Jones (LJ) 12-6 potential:
\begin{equation} \label{ljpot}
  U({{r}_{ij}}) =
    \begin{cases}
      4\varepsilon \left[ {{\left( \frac{\sigma }{{{r}_{ij}}} \right)}^{12}}-{{\left( \frac{\sigma }{{{r}_{ij}}} \right)}^{6}} \right] & \text{if} \,\,\,{{r}_{ij}}\le {{r}_{c}},\\
      0 & \text{if}\,\,\, {{r}_{ij}}>{{r}_{c}}.\\
    \end{cases}       
\end{equation}
Here $r_{ij}$ is the distance between two atoms and $ij$ represents pairwise particles. The energy parameter $\varepsilon$ is $1.67\times{10^{-21}}$ J and the length parameter $\sigma$ is $0.34$ nm. $r_c=5.5 \sigma$ is the cut-off distance, beyond which the interaction vanishes. 

The temperature of this system is kept at $T=85$ K using the Nosé-Hoover thermostat. At this temperature, the mass density is $\rho=1.4\times 10^3$ kg/m\textsuperscript{3} and the number density is $n=0.83/\sigma^3$. The density of the vapor phase is about $(1/400) \rho$ so that the effects of vapor on the film are neglected. The surface tension of liquid is $\gamma = 1.52\times{10^{-2}} $ N/m  and the dynamic viscosity is $\mu=  2.87\times{10^{-4}}$ kg/(ms)\,\citep{zhang2020}. {The time step for all simulations is $0.004\sqrt{\varepsilon /(m\sigma ^2)}$  where $m$ is the atmoic mass of argon.}

The initial dimensions of the liquid film (see figure \ref{fig2}(a)) are chosen as $L_x=313.90$ nm, $ L_y=3.14 $ nm. The height of the film varies for three different cases ($h_0=1.2$ nm, 1.6 nm, and 3.14 nm). Therefore, the film is thin ($ L_x\gg h_0$), and the film is quasi-2D ($ L_x\gg L_y$) to save computational costs. {To prepare the initial configuration of the liquid film, the liquid film slab with the desired size is cut from a periodic box of liquid atoms, which makes the film surface flat initially.}  

Conventionally, the solid wall, serving as the boundary condition for the film, is simulated using real solid atoms (real wall) \citep{willis2009enhanced,zhang2019,zhang2020, zhang2020thermal}. In our previous work \citep{zhang2019,zhang2020, zhang2020thermal}, for a real wall, the solid is Platinum with its isotropic $\langle100\rangle$ surface in contact with the liquid.  The liquid-solid interactions are modelled by the same 12-6 LJ potential with $\sigma_{ls}=0.8\sigma$ for the length parameter and $\varepsilon_{ls} =k\varepsilon$. Conventionally, one may vary the energy parameter $\varepsilon_{ls}$ to obtain different levels of slip length. For example, $b=0.68$ nm using $k=0.65$ while $b=8.8$ nm using $k=0.2$\,\citep{zhang2020thermal}. {The disjoining pressure and contact angles of liquid argon on the solid are also changed in this way.} However, the slip length obtained using the real solid is usually in the range of a few ten nanometers\,\citep{bo2010} in MD simulations, which cannot match the micrometer slip length present in experiments. The origin of strong slip in experiments, as discussed earlier in the Introduction, is usually complicated and cannot be straightforwardly simulated in previous MD simulations using LJ potentials. {The use of a real solid wall with a cutoff-distance limited intermolecular interactions between liquid and solid also means that the disjoining pressure \emph{due to the solid} vanishes when the film height is larger than the cutoff distance \citep{willis2009enhanced,zhang2023capillary}, which is undesired.}

{Here we propose a new approach to allow us to generate any values of slip length simply in MD simulations and allow disjoining pressure to be effective at infinite distances physically. Instead of simulating a real wall, we apply a force to fluid atoms to mimic the fluid-substrate interactions and the force acts as the virtual wall, based on the work of \citep{steele1973physical,barrat1999influence,hadjiconstantinou2021atomistic}. The (total) force $\vec{f}$ for a fluid atom interacting with a face-centered-cubic solid substrate by the LJ potential has been calculated analytically by\,\citet{steele1973physical} and it is adopted here with some modifications (see Appendix for detailed discussions):}
\begin{align}\label{virtual_force}
  & \vec{f}={{f}_{x}}\vec{e_x}+{{f}_{z}}\vec{e_z},\quad \text{with} \nonumber\\
 & {{f}_{x}}=\frac{{{(2\pi )}^{2}}\varepsilon}{100{\ell}_{s}^{3}}\left[ \frac{ {{\sigma }^{12}}}{30}{{\left( \frac{\pi }{{{\ell}_{s}}z} \right)}^{5}}{{K}_{5}}\left( \frac{2\pi }{{{\ell}_{s}}}z \right)-2{{\sigma }^{6}} {{\left( \frac{\pi }{{{\ell}_{s}}z} \right)}^{2}}{{K}_{2}}\left( \frac{2\pi }{{{\ell}_{s}}}z \right) \right]\sin \left( \frac{2\pi }{{{\ell}_{s}}}x \right),\nonumber\\
 & {{f}_{z}}=-\frac{dU_0 \left(z\right)}{dz}
\end{align}
Here $\vec{e}_x$ and $\vec{e}_z$ are unit vectors in the $x$ and $z$ direction. $\ell_s$ is the lattice spacing. The $K_5$ and $K_2$ are the modified Bessel function of the second kind. 
The applied force in the $x$ direction $f_x$ decays rapidly with $z$. $f_x$ is closely related to slip length and its sinusoidal-form force represents the energy corrugation of the solid surface. A smaller lattice spacing $\ell_s$ results in a smooth surface energy distribution and then a larger slip length\,\citep{thompson1990origin,barrat1999influence,hadjiconstantinou2021atomistic}. In our MD simulations, the function $K_5$ and $K_2$ are approximated by $\left(\frac{\pi}{2x}\right)^{1/2}e^{-x}$ for simplicity. {Here $U_0 (z)$ is the total interaction energy exerted on a liquid molecule by the (continuous) substrate. As shown in Appendix, $U_0(z)$ is related to the function of disjoining pressure as $U_0(z)=-\phi(z)/n$. Note that disjoining pressure is a function of film height $h$. One has to replace $h$ with $z$ while the form of the function itself is the same. } 

Here $\phi$ takes the usual form\,\citep{Israelachvili}:
\begin{equation}\label{dj}
\phi(h)=\frac{A}{6\pi h^3}-\frac{M}{h^9},
\end{equation}
where typical values $A=1.7\times 10^{-20}$ J and $M=0.018\varepsilon \sigma^6$ are used. {This form of disjoining pressure $h^{-3}-h^{-9}$ is obtained by integrating the 12-6 LJ potential over the entire substrate assuming the substrate is continuous, as shown in Appendix and \citep{Israelachvili,carey2005disjoining,schick1990liquids,dietrich1988phase,macdowell2014disjoining}. There are many other forms of disjoining potential resulting from different liquid and solid properties \citep{be2003}.}
\begin{figure}
\centering
\includegraphics[width=0.5\linewidth]{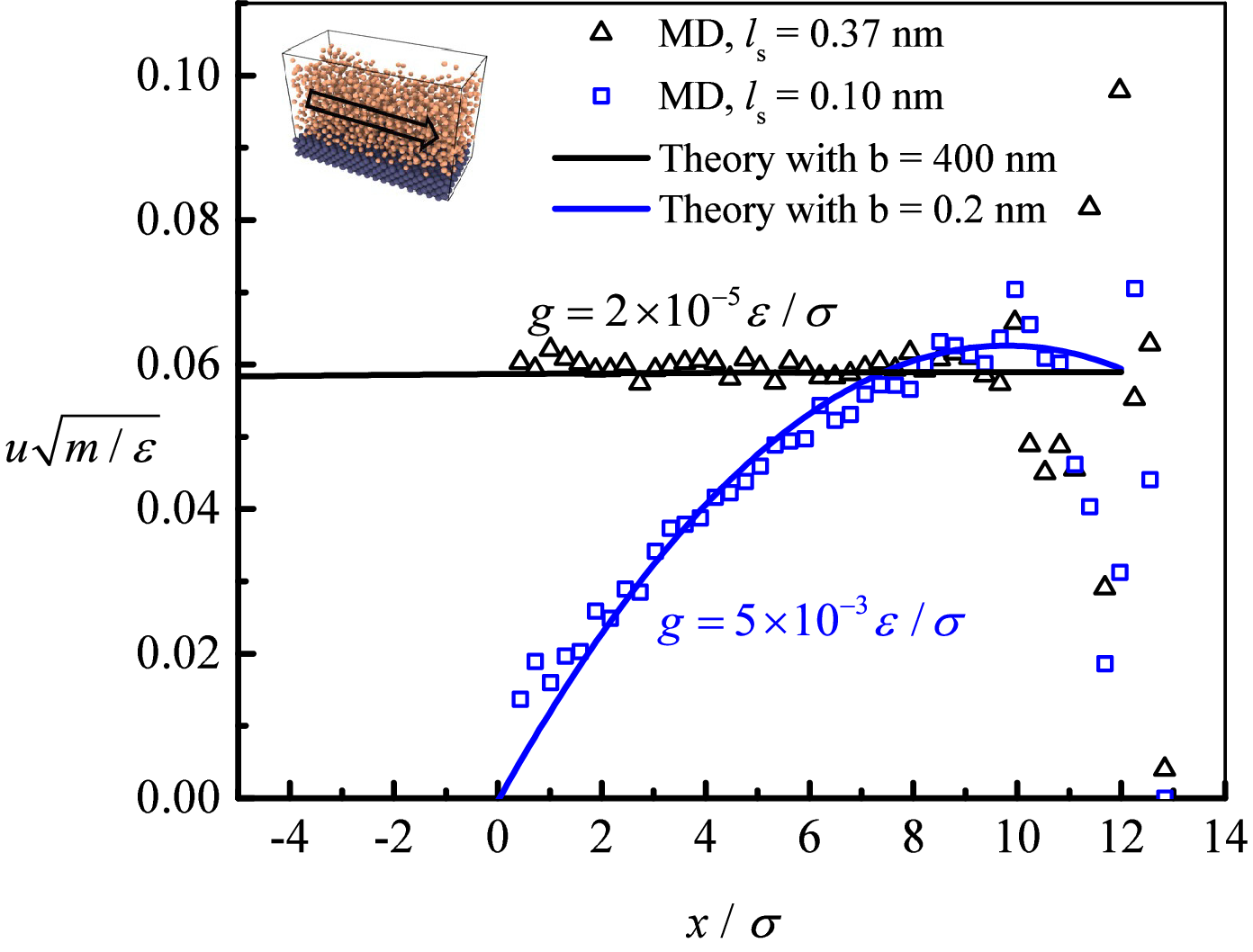}
\caption{Slip length measured using pressure-driven flows past the substrate in molecular simulations. MD results of velocity (symbols) are fitted with analytical solutions (solid lines) to obtain slip length.}
\label{fig3}
\end{figure}

We vary the lattice spacing $\ell_s$ and see what the slip length is from independent simulations where a pressure-driven flow goes past a substrate as shown by the MD snapshot in figure \ref{fig3}. The pressure gradient is created by applying a body force $g$ to the fluid. For $\ell_s=0.37$ nm, one can see that the velocity profile in MD (blue squares in figure \ref{fig3}) is parabolic. However, for $\ell_s=0.10$ nm, the velocity profile in MD (black triangles in figure \ref{fig3}) is nearly constant (plug flow). 

The generated velocity distribution is
\begin{equation}
u(z)=\frac{\rho g}{2\mu }\left[(z-{{z}_{1}})(2{{z}_{2}}-z_1-z)+2b\left(z_2-z_1\right)\right].
\end{equation}
Here $z_1=0$ and $z_2=9.2\sigma$ are the positions of the wall and the free surface respectively. This prediction (solid lines in figure \ref{fig3}) is used to fit the MD results (symbols in figure \ref{fig3}) to obtain the slip length. For $\ell_s=0.37$ nm, a nearly no-slip surface $b=0.2$ nm is achieved. For $\ell_s=0.10$ nm, a strong slip length of $b=400$ nm is obtained. {Note that the choice of thermostats may influence the slip length a little. For example, for water flows inside carbon nanotubes, the Berendsen and Nosé-Hoover thermostats result in very similar slip length, while a smaller slip length is found under the influence of the Langevin thermostat\,\citep{sam2017water}. Thus, the same thermostats should be chosen when measuring the slip length from pressure-driven flows and simulating nanofilm dewetting.}

{To mimic experimental conditions of polymer films on OTS (octadecyltrichlorosilane) and DTS substrates\,\citep{lessel2017nucleated}, where contact angles (disjoining pressure) are about the same for both cases but slip length is very different (no slip on OTS in contrast to 500 nm slip length on DTS), we thus keep the disjoining pressure \eqref{dj} the same for the simulations of weak-slip nanofilm dewetting and strong-slip nanofilm dewetting. { In our simulations, the contact angles of drops after film dewetting for both cases are measured to be about 40 degrees.} This can also facilitate the comparison between weak-slip dewetting and strong-slip dewetting in simulations as the slip length is the only variable. Classically, disjoining pressure decreases with increasing slip length and contact angles. This is, however, far from being universal, as discussed above for the case of polymer films on OTS and DTS substrates. Liquids on hydrophilic substrates with small contact angles can also have large slip length\,\citep{ho2011liquid,rothstein2010slip}. As shown in Appendix, the proposed simulation technique is general and it can include the classic results where slip length and disjoining pressure are coupled and disjoining pressure decreases with increasing slip length and contact angles. As the virtual force is obtained by integrating the LJ potential (especially for $f_x$), this simulation technique is limited for the system where liquid and solid interact with LJ potential.} 

{With the wall modelled by the virtual force, the simulations of nanofilm dewetting are run 2 ns for the case of $h_0=1.2$ nm, 10 ns for $h_0=1.6$ nm, and 100 ns for $h_0=3.14$ nm.}

\section{Results and Discussions}\label{sec5}
\subsection{Evolving spectra of an unstable film with strong slip}
\begin{figure}
\centering
\includegraphics[width=\linewidth]{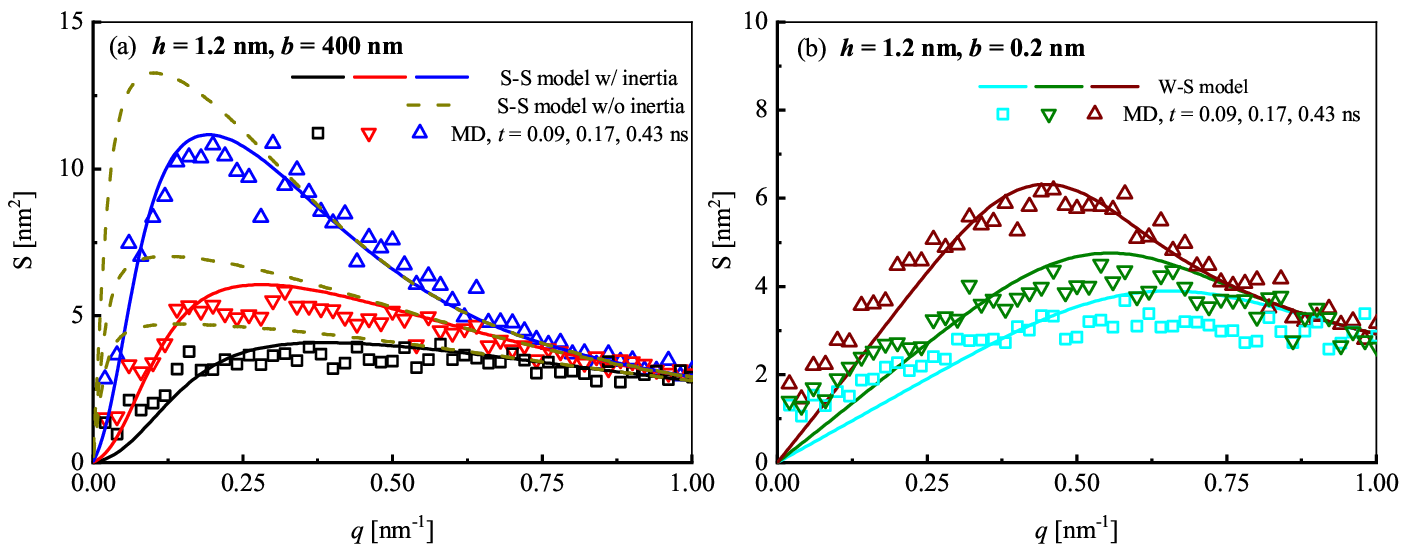}
\captionsetup{justification=justified, singlelinecheck=off} 
\caption{Evolution of the capillary spectra for the film with the thickness $h=1.2$ nm at three different times. (a) the film has a strong slip length $b=400$ nm. Solid lines represent the prediction of the full S-S model while dash lines are the S-S model without the inertial terms.  (b) the film has a weak slip length $b=0.2$ nm.}
\label{fig4}
\end{figure}
As shown in figure \ref{fig2}, a flat film deposited on the substrate experiences the spontaneous growth of perturbations on its surface, leading to film rupture (see figure \ref{fig2}(b)) and droplet formation (see figure \ref{fig2}(c)). To reveal the instability mechanism in the case of the strong-slip dewetting, the evolution of its surface spectra is obtained from MD simulations. 

The instantaneous liquid-vapour interface $h(x,t)$, defined by the usual equimolar surface, is extracted from MD simulations (see \citep{zhang2019} for methods). A discrete Fourier transform of $h(x,t)$ is performed to obtain the amplitude of surface modes $\hat{h}(q,t)$. The surface spectra are thus calculated from the average (root mean square)  of 40 independent simulations. 

The symbols in figure \ref{fig4}(a) represent the MD spectra for the film with thickness $h=1.2$ nm and slip length $b=400$ nm at three different times. Compared to the weak-slip case ($b=0.2$ nm) presented in figure \ref{fig4}(b), one can see that the transient characteristics of the spectra are strongly influenced by the slip length. The spectra for the strong-slip case grow faster than the spectra for the weak-slip case by comparing the blue triangles in figure \ref{fig4}(a) with the wine triangles in figure \ref{fig4}(b). The dominant wavenumber $q_d$ (the one with the maximum amplitude) at different times in the strong-slip dewetting are smaller than those in the weak-slip dewetting (e.g., at $t=0.43$ ns, $q_d\approx 0.25$ nm\textsuperscript{-1} for the strong-slip case in figure \ref{fig4}(a) in contrast to $q_d\approx 0.50$ nm\textsuperscript{-1} for the weak-slip case in figure \ref{fig4}(b)). 

\begin{figure}
\centering
\includegraphics[width=\linewidth]{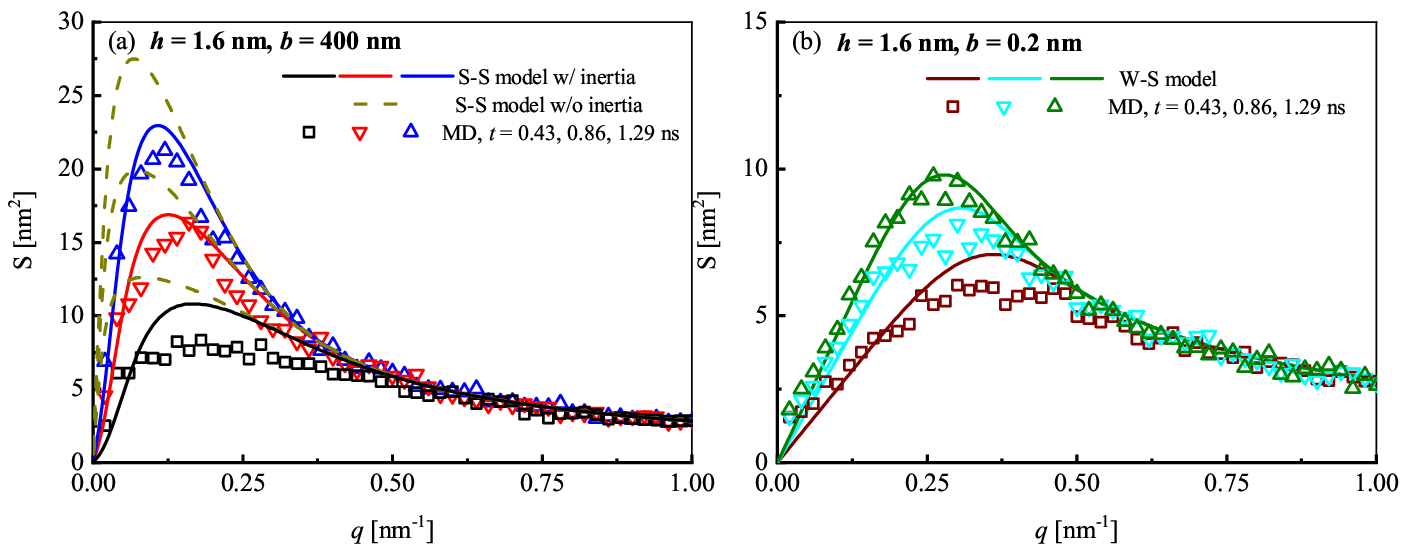}
\captionsetup{justification=justified, singlelinecheck=off} 
\caption{Evolution of the capillary spectra for the film with the thickness $h=1.6$ nm at three different times. (a) the film has a strong slip length $b=400$ nm. (b) the film has a weak slip length $h=0.2$ nm.}
\label{fig5}
\end{figure}

The proposed stochastic models (S-S model and W-S model) and their analytical spectra are used to predict the MD spectra. In our MD simulations, the initial setting of the film surface is flat $\left|\hat{h}(q,0)\right|\approx0$,  so that the deterministic contribution to the surface spectra is $\left|\hat{h}_\mathrm{det}\right|\approx0$. As elaborately investigated in our previous work\,\citep{zhang2019,zhang2020thermal,zhang2020}, the capillary spectra for weak-slip film can be predicted well by the W-S model, e.g., see the solid lines in figure \ref{fig4}(b). However, the W-S model does not apply to the case where the slip is strong, since the W-S model with $b=400$ nm leads to enormous overpredictions compared to MD results (not shown). This is because the derivation of the W-S model requires the slip length on the order of the film thickness. To predict the spectra of the unstable film with strong slip $b=400$ nm, the analytical spectra \eqref{gsss} of the newly derived S-S model is adopted and it agrees excellently with MD results (see the solid lines in figure \ref{fig4}(a)). The symbols in figure \ref{fig5}(a) show the MD spectra for the film with a larger thickness $h=1.6$ nm. Again, our analytical spectra can predict the MD results very well. Note that for large wavenumbers, the spectra at different times simply collapse into the static spectrum $S_s=\sqrt{L_x k_BT/(L_y\gamma q^2)}$.

Another interesting finding is that the assumption of Stokes flow i.e., negligible inertia, breaks down when the value of slip length is increased from a weak slip to a strong slip. Ignoring the inertial terms in the derived S-S SLE (the left-hand-side terms of \eqref{sle12}), the surface spectra are simply
\begin{equation}
S\left( q,t \right)=\sqrt{\frac{{{L}_{x}}}{{{L}_{y}}}\frac{{{k}_{B}}T}{\gamma {{q}^{2}}+d\phi /dh}\left[ 1-{{e}^{-2Dt/C}} \right]},
\end{equation}
which however overpredicts the MD results (see the dash line in figure \ref{fig4}(a) and \ref{fig5}(a)).
Note that the W-S model where inertia is also absent works well for predicting the MD results of weak-slip dewetting. Therefore, it is the strong slip that brings the inertia into effect. In the weak-slip regime, the Reynolds number is
\begin{equation}
\mathrm{Re}_{W-S}=\frac{\rho u_0 h_0}{\mu}=\chi^3 \frac{\rho \gamma h_0}{\mu^2}=\chi^3 La.
\end{equation}
Here we have used the momentum balance in the horizontal direction to estimate the characteristic velocity $u_0\sim \chi^3 \gamma/\mu$ as discussed before. However, in the strong-slip regime, the $\mathrm{Re}_{S-S}$ is 
\begin{equation}
\mathrm{Re}_{S-S}=\frac{\rho u_0 h_0}{\mu}=\chi \frac{\rho \gamma h_0}{\mu^2}=\chi  La,
\end{equation}
based on the momentum balance in the vertical direction. Therefore $\mathrm{Re}_{S-S}$ of the strong-slip case is $1/\chi^2$ larger than the weak-slip case. On the other hand, the Laplace number $\mathrm{La}$ in our simulations is calculated to be $0.3$ which is consistent with the assumption of deriving the strong-slip model where the $\mathrm{La}\sim 1$. Both conditions (strong slip and $\mathrm{La}=0.3$) make inertia non-negligible in our simulations.
\begin{figure}
\includegraphics[width=\linewidth]{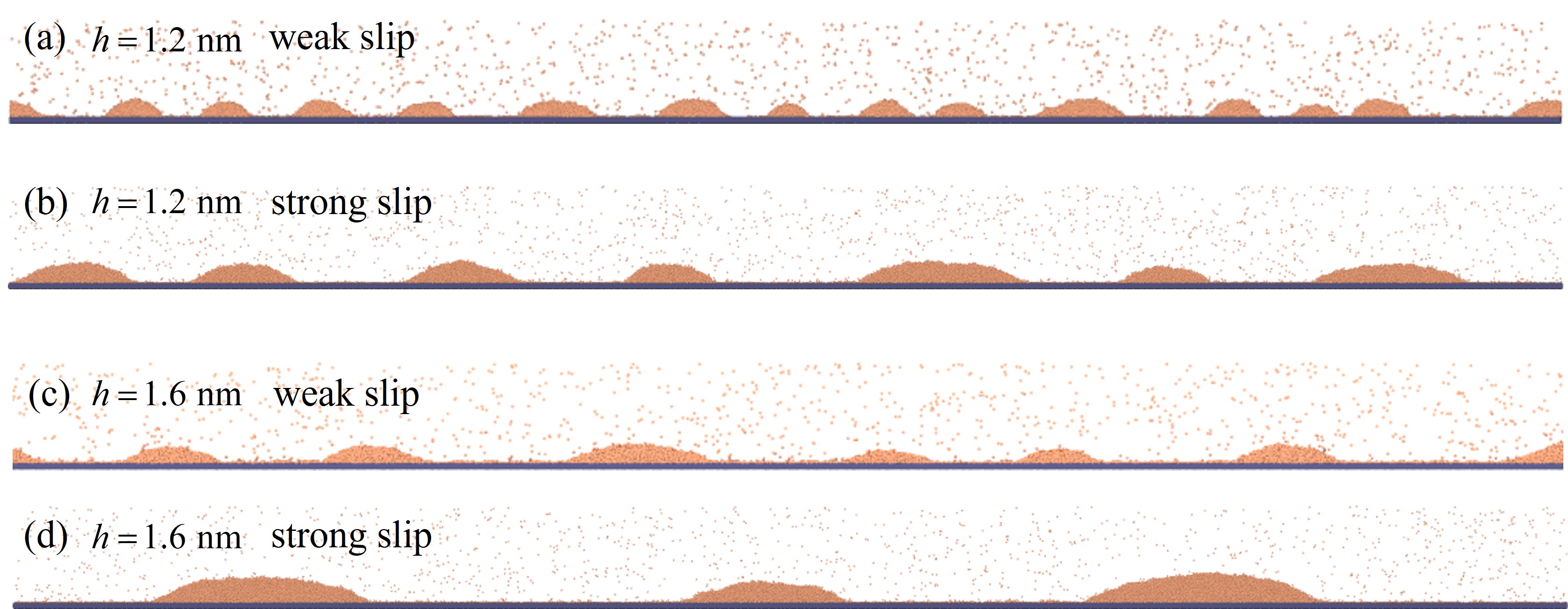}
\caption{Droplets form after the rupture of the film. (a) for the film with $h=1.2$ nm and $b=0.2$ nm. (b) for the film with $h=1.2$ nm and $b=400$ nm. (c) for the film with $h=1.6$ nm and $b=0.2$ nm. (d) for the film with $h=1.6$ nm and $b=400$ nm.}
\label{fig6}
\end{figure}
\subsection{Number of droplets}
\begin{figure}
\centering
\includegraphics[width=0.9\linewidth]{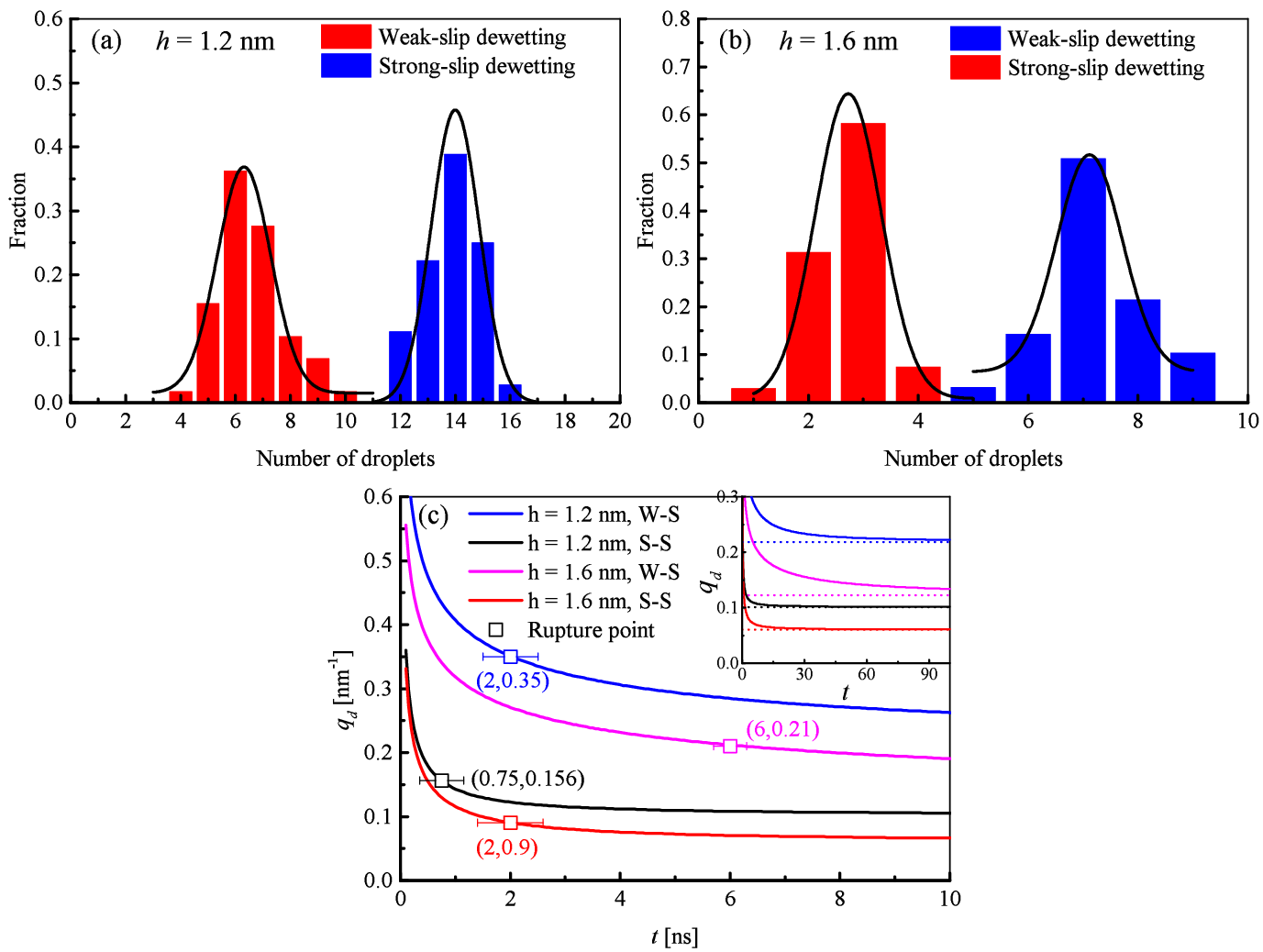}
\caption{(a) Distribution of the number of droplets for the film with $h=1.2$ nm. (b) Distribution of the number of droplets for the film with $h=1.6$ nm. (c) The decrease of the dominant wavenumber with time predicted by the \eqref{gsss} (solid lines) to its asymptotic value by the deterministic lubrication equation (dash lines in the inset). The squares represent the dominant wavenumber at the time of film rupture.}
\label{fig7}
\end{figure}
As shown in figure \ref{fig6}, one distinct feature of the strong-slip dewetting (see figures \ref{fig6}(b) and (d)) is having fewer droplets after the film ruptures compared to the weak-slip dewetting (see figures \ref{fig6}(a) and (c)). For each case with a specific film thickness and slip length (for example, in figure \ref{fig6}(a), $h=1.2$ nm and $b=0.2$ nm), about 40 independent simulations have been run and the probability distribution of the number of formed droplets is calculated and shown in figures \ref{fig7}(a) for $h=1.2$ nm and \ref{fig7}(b) for $h=1.6$ nm. 

In terms of the thinner film $h=1.2$ nm, the average number of droplets for weak-slip dewetting is $N_a=14$ (see the blue bars in figure \ref{fig7}(a) and a typical MD snapshot in figure \ref{fig6}(a)), while the number is $N_b=6.5$ for the strong-slip dewetting (see the red bars in figure \ref{fig7}(a) and a typical MD snapshot in figure \ref{fig6}(b)). For the thicker film with $h=1.6$ nm, the mean number of droplets for weak-slip dewetting is $N_c=7$ (see the blue bars in figure \ref{fig7}(b) and the MD snapshot in figure \ref{fig6}(c)) while the number is $N_d=3$ for strong-slip dewetting (see the red bars in figure \ref{fig7}(b) and the MD snapshot in figure \ref{fig6}(d)).

The number of droplets is connected with the dominant wavenumber $q_d$ (or the dominant wavelength $\lambda_d=2\pi/q_d$). In the deterministic situation, the dominant wavenumber is constant over time. For the deterministic W-S equation, it is $q_{d}=\sqrt{-({d\phi}/{dh})/({2\gamma})}\approx \sqrt{{A}/({4\pi\gamma h_0^4})}$,
while for the deterministic S-S equation, it is the solution when $d\omega_1(q)/dt=0$.

However, as found from MD simulations shown in figure \ref{fig4} and figure \ref{fig5}, the dominant wavenumber decreases with time, which is surely beyond the prediction from the deterministic lubrication equation. Based on the analytical spectra of the W-S model \eqref{wsss} and S-S model \eqref{gsss}, figure \ref{fig7} (c) shows the theoretical prediction of the evolving dominant wavenumber. One can find that thermal fluctuations induce a dominant wavenumber (see the solid lines) much higher than its corresponding deterministic prediction (see the dash lines in the inset) and it only gradually decreases to the deterministic predictions over time. In our simulations, the film can rupture far before the dominant wavenumber reaches its asymptotic value.

Therefore, we measure the average time of rupture $t_r$ from MD simulations (shown by the symbols in figure \ref{fig7}(c)) and use the analytical theory to predict the dominant wavenumber at the time of rupture. For $h=1.2$ nm and $b=0.2$ nm, the rupture time  
is about $t_r=2$ ns and the dominant wavenumber at this time is $q_d=0.35$ nm\textsuperscript{-1}. The dominant wavenumber corresponds to $ q_d L_x/(2\pi)\approx 17$ droplets, in agreement with the $14$ measured directly from MD simulations. For $h=1.2$ nm and $b=400$ nm, the rupture is much faster $t_r=0.75$ ns and the dominant wavenumber at this time is $q_d=0.156$ nm\textsuperscript{-1}. This dominant wavenumber translates into $7.8$ droplets, in agreement with the $6.5$ measured directly from MD simulations. Therefore increasing slip length from weak slip to strong slip can indeed reduce the number of droplets by decreasing the dominant wavenumber. 

Since the strong-slip dewetting and the weak-slip dewetting starts with the same surface condition (flat surface), the deviations for both cases in the dominant wavenumber begin at a very early time (see figure \ref{fig7}(c)). The smaller dominant wavenumber in the strong-slip dewetting, compared with the weak-slip dewetting, becomes the obvious feature of the strong-slip spectra shown in figure \ref{fig4}(a).

By increasing the film thickness from $h=1.2$ nm to $h=1.6$ nm, the number of droplets is also decreased. For example, for the strong-slip dewetting, the average number of droplets changes from 6.5 to 3 by with increased film thickness as shown in figure \ref{fig7}(b). This can be also predicted by the dominant wavenumbers shown in figure \ref{fig7}(c) where the dominant wavenumber is smaller for a larger film thickness.
\subsection{Surface roughening with strong slip}
\begin{figure}
\centering
\includegraphics[width=0.5\linewidth]{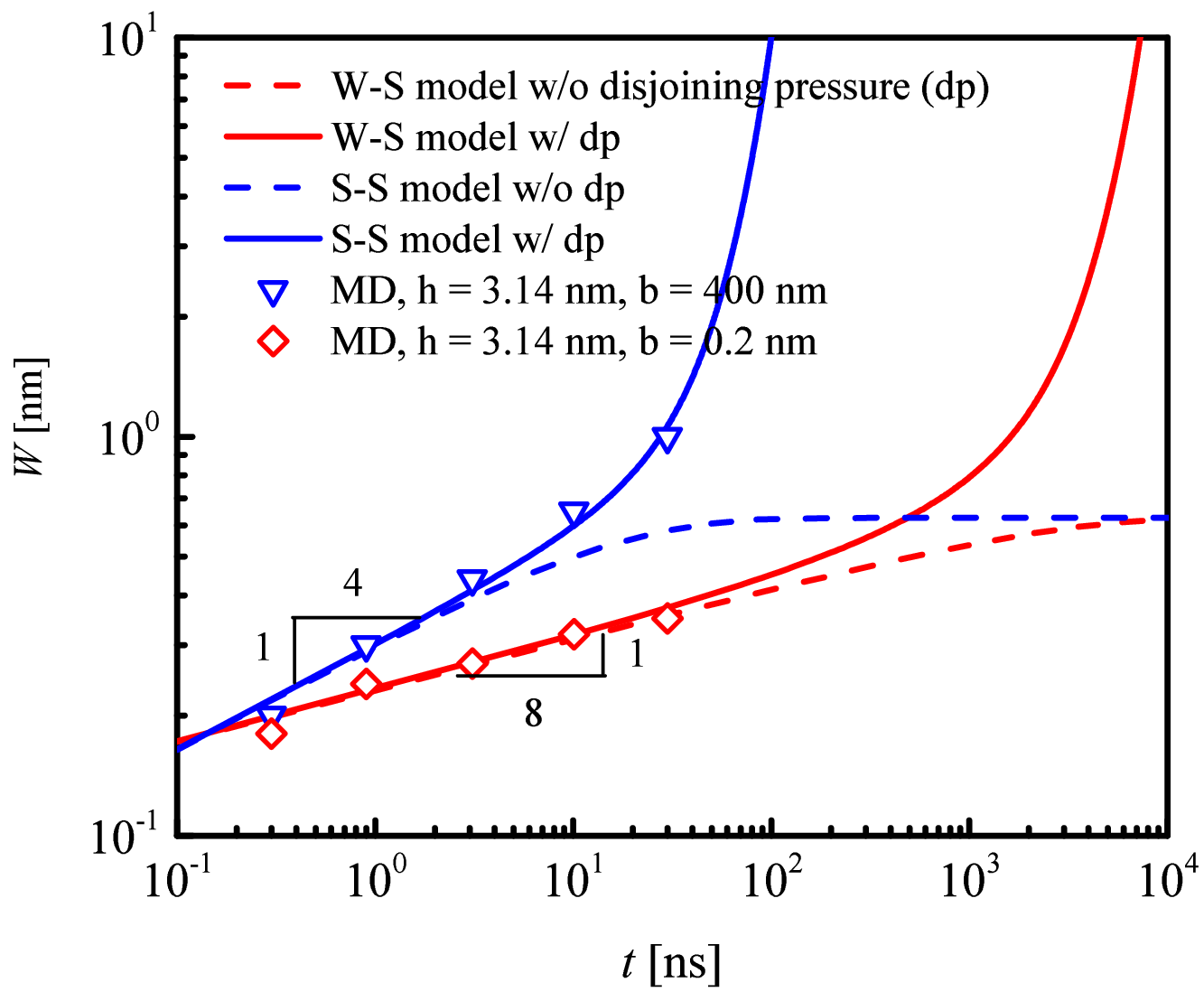}
\caption{The growth of roughness on films with weak slip and strong slip.}
\label{fig8}
\end{figure} 
The growth of spectra is equivalent to the surface roughening of the free surface based on Parseval's theorem:
\begin{equation}\label{ws}
W(t)=\sqrt{\frac{1}{{{L}_{x}}}\left\langle\int_{0}^{{{L}_{x}}}{{{\left( \delta h \right)}^{2}}\,dx}\right\rangle}=\sqrt{\frac{1}{2\pi {{L}_{x}}}\int_{\frac{2\pi}{{{L}_{x}}}}^{\frac{2\pi K}{{{L}_{x}}}}{S^2\, dq}}\, ,
\end{equation}
where $W(t)=\sqrt{\overline{h^2}-{\overline{h}}^2}$ is the root-mean-square roughness of the free surface. $K$ is the number of bins used to extract the surface profile from MD simulations, which provides an upper bound on the wavenumbers that can be extracted\,\citep{zhang2020thermal}. 

In our study, the roughening of a flat liquid surface is due to thermal fluctuations (and also disjoining pressure). Surface roughening or growth resulting from other forms of randomness is a common occurrence in nature. Examples include the propagation of wetting fronts in porous media, the growth of bacterial colonies, and the atomic deposition process in the manufacture of computer chips\,\citep{ba1995}. These evolutions of the profile of a growing interface can be described by stochastic partial differential equations (SPDE). One of the most famous examples is the Kardar–Parisi–Zhang (KPZ) equation\,\citep{kardar1986dynamic}. Scaling relations for surface roughening are then used to analyze the SPDE, which can be summarized as\,\citep{ba1995} 
\begin{equation}
 W\left(t\right)\sim L^{\alpha} f(t/L^m),
\end{equation}
where $L$ is the system size, $f(v)=v^{\kappa}$ for $v\ll 1$ (during roughness growth), and $f(v)=1$ for $v\gg 1$ (at roughness saturation). The time to transition, between roughness growth and saturation, scales with $t_s \sim L^m$. The three exponents ($\alpha$, $m$ and $\kappa$) define a universality class, and are here related by $\kappa = \alpha /m$.
For example, for the one-dimensional KPZ equation, $(\alpha=1/2,m=1/3,\kappa=3/2)$. Basically, the roughness will grow as a power law of time $W\sim t^{\kappa}$ before getting saturated, which is a result of the balance of the deterministic forces (such as surface tension in our study) and stochastic forces (such as thermal motions of molecules in our study). 

In our previous work\,\citep{zhang2020thermal}, we have shown that for a weak-slip film roughened by thermal fluctuations (with negligible effects of disjoining pressure), a university class $(\alpha=1/2,m=4,\kappa=1/8)$ exists and $W\sim t^{1/8}$ before the roughness saturation. It is interesting to see how strong slip changes the scaling of surface roughening. We note that the existence of the disjoining pressure breaks the potential balance of surface tension and thermal fluctuations, resulting in the unbounded growth of the surface roughness. However, the initial growth of surface roughness, when the disjoining pressure is weak, may still be described by scaling laws.

As shown in figure \ref{fig8}, we calculate the surface roughening of a film with the thickness $h=3.14$ nm for different levels of slip length. In terms of the weak-slip film $b=0.2$ nm and without disjoining pressure, the roughness grows with $W\sim t^{1/8}$ and then becomes saturated eventually (see the blue dash line in figure \ref{fig8}). When disjoining pressure is considered, the growth of the roughness becomes unbounded (see the red solid line). However, the initial stage of the growth of roughness ($t<100$ ns) is unaffected by disjoining pressure so that the $W\sim t^{1/8}$ is still valid.

For the strong-slip film $b=400$ nm and without disjoining pressure, we find $W\sim t^{1/4}$ as shown by the blue dash line in figure \ref{fig8}, which can be derived as follows. $\alpha$ can be obtained by considering the surface at saturation, i.e. from the static spectrum given by $S_s=\sqrt{L_x k_BT/(L_y\gamma q^2)}$. Substituting the static spectrum into \eqref{ws} leads to
\begin{equation}
W_s=\sqrt{\frac{1}{2\pi L_y}\frac{k_BT}{\gamma}\left(\frac{L_x}{2\pi}-\frac{L_x}{2\pi K}\right)}.
\end{equation}
For large $K$, which is the case in our MD simulations, $W_s$ becomes independent of $K$, and one can find that $W_s=\sqrt{\frac{L_x}{4\pi^2 L_y}\frac{k_BT}{\gamma}}\sim {L_x}^{{1}/{2}}$.

An upper estimate on the transition time $t_s\sim L^m$, between growth and saturation, can be estimated from the inverse of the dispersion relation at the largest permissible wavelength ($q={2\pi}/{L_x}$). Examining the dispersion relation indicates that there are three time scales for the spectra to reach thermal equilibrium (without disjoining pressure) whose maximum is the right time scale, namely,
 \begin{align}
  t_s=&\frac{1}{\mathrm{min} \left(|2\omega_1\left(q=2\pi/L_x\right)|, |2\omega_2\left(q=2\pi/L_x\right)|, C \left(q=2\pi/L_x\right)\right)}\nonumber\\
  =&\frac{1}{|2\omega_1\left(q=2\pi/L_x\right)|}\nonumber\\
 \sim&{L_x^2},
 \end{align}
where we have used the condition $C=\frac{\mu }{\rho }\left( 4{{q}^{2}}+\frac{1}{{{h}_{0}}b} \right)\approx \frac{4{{q}^{2}}\mu }{\rho } $ due to the fact that $4q^2h_0b=4/\chi\gg1$ (using the long-wave condition $qh_0 \sim \chi$ and the strong-slip condition $bh_0 \sim \chi^{-2}$).

Therefore we find the power $m=2$ and $\kappa=\alpha/m=1/4$ for the strong-slip case. In summary, the university class $(\alpha=1/2,m=2,\kappa=1/4)$ is found for the surface growth of a thin film with strong slip and negligible disjoining pressure. When disjoining pressure is considered, the initial growth of the surface may still be described by $W\sim t^{{1}/{4}}$ (as shown by the blue solid line in figure \ref{fig8}), which is much faster than the $W\sim t^{{1}/{8}}$ of the weak-slip case. These scalings are confirmed in MD simulations (see the symbols in figure \ref{fig8}). 
\section{Conclusions}\label{sec6}
In this work, the instability of molecularly thin liquid films on substrates with strong slip is investigated using both molecular dynamics simulations and a newly developed stochastic lubrication equation.
A new method is proposed to generate strong slip length in molecular simulations. Our simulations reveal that the strong-slip dewetting is much faster than the weak-slip dewetting and has fewer droplets formed compared with the weak-slip dewetting. Using a long-wave approximation to the equations of fluctuating hydrodynamics, a new stochastic lubrication equation considering strong slip and inertia effects is derived.
By the linear stability analysis, the analytical surface spectrum is obtained and it can explain the differences between strong-slip dewetting and weak-slip dewetting observed in simulations. We further find that inertial effects, which are usually ignored in weak-slip dewetting, come into play in the strong-slip dewetting due to the enhanced velocity by the strong slip. 

When the effect of disjoining pressure is negligible, the derived strong-slip stochastic lubrication equation possesses a university class $(1/2,2,1/4)$ in contrast to the $(1/2,4,1/8)$ of the weak-slip stochastic lubrication equation. In our simulations, the effect of disjoining pressure is important. However, the surface roughening of strong-slip dewetting at the initial stage may be still described by $W\sim t^{1/4}$ in contrast to the $W\sim t^{1/8}$ for weak-slip dewetting.

Experiments of the rupture of polymer nanofilms on \emph{no-slip} SiO\textsubscript{2}-coated silicon wafers have demonstrated the great effects of thermal fluctuations\,\citep{fe2007}. The experimental surface spectra can be predicted with the \emph{no-slip} film stochastic lubrication equation\,\citep{gr2006,fe2007}. However, polymer films on DTS solid can have slip length up to several micrometers\,\citep{baumchen2009slip}. Therefore, experiments on the rupture of polymer nanofilms on DTS solid can be used to validate the newly developed theory here. However, care should be taken to make sure the dewetting of polymer films is initiated by disjoining pressure since polymer films can be easily contaminated (leading to nucleation dewetting)\,\citep{jacobs1998thin}. As polymers have a very large viscosity $\sim 10^4$ kg/(ms), the Laplace number is very small so that the effects of inertia can be safely neglected. For metallic films, however, its viscosity is about $\sim 10^{-3}$ kg/(ms) so that inertial effects can be important\,\citep{di2016,gonzalez2016inertial}.  

{In the limit of infinite slip, the derived strong-slip stochastic lubrication equation leads to a new model for free films, which can be readily used to study, for example, the symmetrical breakup of a foam film under the effects of thermal fluctuations\,\citep{erneux1993nonlinear,vaynblat2001rupture}. In this work, we focus on the linear stability theory and molecular simulations of nanofilm dewetting, numerical solutions to these stochastic lubrication equations remain to be investigated in the future, e.g, see \citep{gr2006,ne2015,di2016,du2019,sh2019,zhao2020dynamics}.} Future directions may also include the effects of contamination (such as surfactants)\,\citep{zhang2023capillary} and evaporation\,\citep{or1997,craster2009dynamics} on film stability which often occur at small scales.
\section*{Acknowledgement}
We are grateful for the discussions with Detlef Lohse, Duncan Lockerby, James Sprittles, and Vincent Bertin. This work is supported by NWO under the project of ECCM KICKstart DE-NL 20002799. 
\section*{Declaration of interests}
The authors report no conflict of interest.
\renewcommand{\theequation}{A\arabic{equation}}
\setcounter{equation}{0}
\section*{Appendix}
In this Appendix, the process of obtaining $f_x$ and $f_z$ is explained based on the previous work in literature, e.g., \citet{steele1973physical,barrat1999influence,hadjiconstantinou2021atomistic}. First, the molecular mechanism of disjoining pressure is reviewed briefly. For the 12-6 Leonard Jones potential between fluid atoms and solid atoms as shown by \eqref{ljpot}, the total interaction energy exerted on a liquid molecule by a semi-infinitely extended and continuous substrate with density $n_s$ is (see p209 of \citet{Israelachvili})
\begin{align}\label{utot_dis}
  & {{U}_{tot}}(D)=\int_{z=D}^{z=\infty }{\int_{x=0}^{x=\infty }{\left( \frac{\alpha_u }{{{r}^{12}}}-\frac{\beta_u }{{{r}^{6}}} \right)}}2\pi x{{n}_{s}}dxdz  \nonumber\\ 
 & =\int_{z=D}^{z=\infty }{dz\int_{x=0}^{x=\infty }{\left[ \frac{\alpha_u }{{{\left( {{z}^{2}}+{{x}^{2}} \right)}^{6}}}-\frac{\beta_u }{{{\left( {{z}^{2}}+{{x}^{2}} \right)}^{3}}} \right]}}2\pi x{{n}_{s}}dxdz  \nonumber\\ 
 & =\frac{\pi \alpha_u {{n}_{s}}}{45{{D}^{9}}}-\frac{\pi \beta_u {{n}_{s}}}{6{{D}^{3}}}. 
\end{align}
Here $D$ is the distance of the fluid atom to the solid surface. We use $\alpha_u$ and $\beta_u$ for a more general case and one can have $\alpha_u =4\varepsilon {{\sigma }^{12}}$  and $\beta_u =4\varepsilon {{\sigma }^{6}}$ as \eqref{ljpot}. One can see that the gradient of this potential is a force that results in pressure variations inside the liquid such that $\nabla p=-n_l \nabla U_{tot}$ \citep{carey2005disjoining}. Therefore, the pressure inside the liquid is
\begin{equation}
p\left( D \right)\equiv -{{n}_{l}}{{U}_{tot}}=\frac{\pi \beta_u {{n}_{s}}{{n}_{l}}}{6{{D}^{3}}}-\frac{\pi \alpha_u {{n}_{s}}{{n}_{l}}}{45{{D}^{9}}}.
\end{equation}
The disjoining pressure $\phi(h)$ on the film surface is then $p(D)$ evaluated at $h$:
\begin{equation} \label{eqa3}
\phi \left( h \right)\equiv p(h)\equiv -{{n}_{l}}{{U}_{tot}}\left( h \right)=\frac{\pi \beta {{n}_{s}}{{n}_{l}}}{6{{h}^{3}}}-\frac{\pi \alpha {{n}_{s}}{{n}_{l}}}{45{{h}^{9}}},
\end{equation}
which is the $h^{-3}-h^{-9}$  kind of disjoining pressure used in this work with different perfectors there (see \eqref{dj}) to account for contact angles close to the experimental values. {Though \eqref{eqa3} is widely used, especially in the field of fluid dynamics, one has to keep in mind that it is an approximation (though being good) of the real disjoining pressure in the liquid. This is because its derivations ignore the structures of liquid-vapor interfaces or liquid-solid interfaces where the liquid density is not simply uniform \citep{schick1990liquids}. In fact, the liquid-vapor interface may contribute additional disjoining pressure in the liquid, though it does not necessarily affect the film instability.} We stress that there are two findings here. First, the cutoff distance are often used in molecular simulations to reduce computational costs. By doing this, the solid does not cause disjoining pressure at the liquid-vapor interface when the film thickness is larger than the cutoff distance as there are no interactions between the film surface and the solid substrate. Secondly, treating the solid as a \emph{continuous} material with density $n_s$, which is the usual way to obtain disjoining pressure from intermolecular potentials as shown above, actually eliminates the drag force from the wall to the liquid in the direction parallel to the substrate, leading to  \emph {free slip}. Therefore, treating the solid as a discontinuous crystal to obtain the correct force, especially in the $x$ direction, is necessary. This was done by \citet{steele1973physical}. The total potential for a fluid particle (gas in \citet{steele1973physical}) above a multilayer solid crystal is
\begin{align}\label{utot_steele}
  & {{U}_{tot}}=\frac{2\pi {{\varepsilon }_{gs}}}{{{a}_{s}}}\sum\limits_{\alpha }{\left\{ q\left( \frac{2}{5}\frac{\sigma _{gs}^{12}}{z_{\alpha }^{10}}-\frac{\sigma _{gs}^{6}}{z_{\alpha }^{4}} \right) \right.}+... \nonumber\\ 
 & \left. +\sum\limits_{g\ne 0}{\sum\limits_{k=1}^{q}{\exp (i\vec{g}\cdot \left[ {{{\vec{m}}}_{k}}+\vec{\tau } \right])\times \left[ \frac{\sigma _{gs}^{12}}{30}{{\left( \frac{g}{2{{z}_{\alpha }}} \right)}^{5}}{{K}_{5}}\left( g{{z}_{\alpha }} \right)-2\sigma _{gs}^{6}{{\left( \frac{g}{2{{z}_{\alpha }}} \right)}^{2}}{{K}_{2}}\left( g{{z}_{\alpha }} \right) \right]}} \right\}. 
\end{align}
Here $a_s=\ell_s^2$ is the area of the unit lattice cell with a lattice spacing $\ell_s$. $\alpha$ is the layer number of crystal layers and $z_{\alpha}$ is its location. $\vec{\tau}$ is the two-dimensional translation vector and $\vec{g}$ is a multiple of the reciprocal lattice vectors. $\vec{m}$ is a vector that gives the location of the $k_{th}$ atom in the unit cell.
Basically, $U_{tot}$ in \eqref{utot_steele} can be split into two parts $U_0(z)$ and $U_1(xyz)$ (also see \citet{barrat1999influence}):
\begin{align}
  & {{U}_{tot}}\left( xyz \right)={{U}_{0}}(z)+{{U}_{1}}(xyz), \nonumber\\ 
 & {{U}_{0}}(z)=\frac{2\pi }{{{a}_{s}}}\sum\limits_{\alpha }{\left\{ q\left( \frac{2}{5}\frac{\sigma _{gs}^{12}}{z_{\alpha }^{10}}-\frac{\sigma _{gs}^{6}}{z_{\alpha }^{4}} \right) \right\}}, \nonumber\\ 
 & {{U}_{1}}(xyz)=\frac{2\pi {{\chi }_{gs}}}{{{a}_{s}}}\sum\limits_{g\ne 0}{\sum\limits_{k=1}^{q}{\exp (i\vec{g}\cdot \left[ {{{\vec{m}}}_{k}}+\vec{\tau } \right]) \left[ \frac{\sigma _{gs}^{12}}{30}{{\left( \frac{g}{2{{z}_{\alpha }}} \right)}^{5}}{{K}_{5}}\left( g{{z}_{\alpha }} \right)-2\sigma _{gs}^{6}{{\left( \frac{g}{2{{z}_{\alpha }}} \right)}^{2}}{{K}_{2}}\left( g{{z}_{\alpha }} \right) \right]}} 
\end{align}
$U_0(z)$ in the limit of a continuous solid with number density $n_s$ is ${{U}_{0}}=\frac{2\pi {{n}_{s}}{{\varepsilon }_{gs}}}{3}\left[ \frac{2}{15}\frac{\sigma _{gs}^{12}}{{{z}^{9}}}-\frac{\sigma _{gs}^{6}}{{{z}^{3}}} \right]$, which is the same as \eqref{utot_dis}. In terms of $U_1(xyz)$, considering that only the potential from the first lattice layer and the shortest reciprocal lattice vectors is the most important \citep{barrat1999influence,hadjiconstantinou2021atomistic}, it can be simplified to
\begin{equation}
{{U}_{1}}(xyz)=\frac{2\pi {{\varepsilon}_{gs}}}{{{a}_{s}}}\left[ \frac{\sigma _{gs}^{12}}{30}{{\left( \frac{\pi}{\ell_s z} \right)}^{5}}{{K}_{5}}\left(  \frac{2\pi }{{{\ell }_{s}}}z \right)-2\sigma _{gs}^{6}{{\left( \frac{\pi}{\ell_s z} \right)}^{2}}{{K}_{2}}\left(  \frac{2\pi }{{{\ell }_{s}}}z \right) \right]\left[ \cos \left( \frac{2\pi }{{{\ell }_{s}}}x \right)+\cos \left( \frac{2\pi }{{{\ell }_{s}}}y \right) \right]
\end{equation}
Finally, the potential exerted by the wall to each fluid atom that we used in the work is (the $y$ direction may be ignored if the simulation is (quasi) two dimensional)
\begin{align}\label{A7}
  & {{U}_{tot}}\left( xz \right)=\frac{2\pi {{\varepsilon }_{gs}}}{3}\left[ \frac{2}{15}\frac{\sigma _{gs}^{12}}{{{z}^{9}}}{{C}_{1}}-\frac{\sigma _{gs}^{6}}{{{z}^{3}}}{{C}_{2}} \right]+... \nonumber \\ 
 & +{{C}_{3}}\frac{2\pi {{\varepsilon}_{gs}}}{{{a}_{s}}}\left[ \frac{\sigma _{gs}^{12}}{30}{{\left( \frac{\pi}{\ell_s z} \right)}^{5}}{{K}_{5}}\left(  \frac{2\pi }{{{\ell }_{s}}}z \right)-2\sigma _{gs}^{6}{{\left( \frac{\pi}{\ell_s z} \right)}^{2}}{{K}_{2}}\left(  \frac{2\pi }{{{\ell }_{s}}}z \right) \right] \cos \left( \frac{2\pi }{{{\ell }_{s}}}x \right) , 
\end{align}
where we have added factors $C_1,C_2$, and $C_3$ so that they can be tuned ($C_3=1/100$ in this work) for the slip length, disjoining pressure, and contact angles reported in literature. More generally, we can use the function of disjoining pressure to replace the first term in \eqref{A7}, namely, the total potential ${{U}_{0}}=-\frac{\phi \left( z \right)}{{{n}_{l}}}$. The force expressed in \eqref{virtual_force} is thus ${f_x}=-d U_1(xz) /dx$ (also see \citet{barrat1999influence} and ${f_z}=-d U_0 /dz$ . If one wants that the slip length and disjoining pressure to be coupled, one can use \eqref{A7} with $C_1=C_2=C_3=1$ where slip length increases with contact angles but disjoining pressure decreases with increasing slip length.).  
\bibliographystyle{jfm}
\bibliography{reference}

\end{document}